\pdfoutput=1
\documentclass[11pt,a4paper]{article}
\usepackage[margin=1in]{geometry}
\usepackage[T1]{fontenc}
\usepackage{lmodern}
\usepackage{textcomp}
\usepackage{amsmath,amssymb}
\usepackage{graphicx}
\usepackage[labelfont=bf,font=small,justification=justified,singlelinecheck=false]{caption}
\usepackage{booktabs}
\usepackage{tabularx}
\usepackage{array}
\usepackage{xcolor}
\usepackage{url}
\usepackage[super,sort&compress]{natbib}
\setcitestyle{citesep={,}}
\usepackage[colorlinks=true,linkcolor=blue!50!black,citecolor=blue!50!black,urlcolor=blue!50!black]{hyperref}
\usepackage{setspace}
\DeclareCaptionLabelSeparator{bar}{ | }
\renewcommand{\figurename}{Fig.}
\newcommand{\model}{Rescene-ESM}
\newcommand{\co}{CO$_2$}
\newcommand{\wm}{W\,m$^{-2}$}
\newcommand{\ED}[1]{Extended Data Fig.~#1}
\newcommand{\SI}[1]{Supplementary Fig.~#1}
\newcommand{\ST}[1]{Supplementary Table~#1}

\title{\textbf{\model: an Earth system model built around\\ a frozen AI weather model}}
\author{
Minjong Cheon$^{1,*}$ \quad Kaoru Kakinuma$^{2}$\\[4pt]
\small\textit{$^{1}$Department of Computer Science, Korea National Open University, Seoul, Republic of Korea}\\
\small\textit{$^{2}$Frontier Research Institute for Interdisciplinary Sciences, Tohoku University, Sendai, Japan}\\
\small $^{*}$\href{mailto:jmj2316@mail.knou.ac.kr}{jmj2316@mail.knou.ac.kr} \quad ORCID 0000-0001-7589-4078
}
\date{November 2026}

\begin{document}

\maketitle

\begin{abstract}
Earth system models underpin climate projections, but their cost limits how many scenarios, ensemble members and
impact pathways can be explored. Machine-learning weather models forecast as
skillfully as physics-based models at a fraction of the cost, yet
coupled to an ocean they have produced weak El Ni\~no variability or have learned it from climate-model
output, and none carries the climate
through rivers to the people exposed to it. Here we present \model, an Earth system model whose coupler, YAHO, yokes a frozen
20.8-million-parameter AI weather model to physically formulated ocean, sea-ice, land, river,
carbon and energy components constrained by observations. Adding the couplings one at a time shows what each contributes:
over a climatological ocean the AI atmosphere varies from year to year by a third of the observed amount over the
oceans, and an interactive mixed layer, an equatorial wave ocean, the residual of the observed sea surface temperature
(SST) equation as stochastic forcing and the land's moisture flux raise this step by step and bring El Ni\~no and its
teleconnections, leaving the mean climate almost unchanged. The full model simulates about 1,000 years per day on
one graphics processing unit, holds the observed climate for a thousand years without drift, reproduces the observed 1979--2022 global warming, carries forced change with a physically based structure around the frozen network (transient climate response
1.60\,K), routes runoff to
212 river gauges with a median seasonal-cycle correlation of 0.79 and, driven by emissions, separates the heatwave
exposure of an SSP1 and an SSP3 world by a factor of 4.7 in 2081--2100, three-quarters of it through the warmer
climate.
\end{abstract}

\section{Introduction}
Projections of climate change and its impacts rest on coupled Earth system models (ESMs), which simulate the atmosphere,
ocean, sea ice, land and carbon cycle together and are compared in coordinated experiments\cite{eyring2016cmip6}. Their
cost is high: CMIP6-class models simulate a few to a few tens of model years per wall-clock day on hundreds to thousands
of processor cores\cite{balaji2017cpmip}. Internal variability is large enough that forced changes in regional climate,
and in extremes in particular, can only be separated from noise with ensembles of tens of
members\cite{deser2020large,lehner2020partitioning}, and every additional emissions or land-use pathway multiplies that
cost. Impact studies add further needs: heat, drought and river floods matter where people live, so projections must
carry the land's water to rivers and the climate to the population exposed to it\cite{jones2015exposure,romanello2023lancet}.

Machine-learning weather models offer a way out of this cost. Trained on the ERA5 reanalysis\cite{hersbach2020era5} to
step the atmosphere forward by a few hours, FourCastNet\cite{pathak2022fourcastnet}, Pangu-Weather\cite{bi2023pangu},
GraphCast\cite{lam2023graphcast}, Stormer\cite{nguyen2024stormer} and Sonny\cite{cheon2026sonny} match or exceed
operational numerical weather prediction in the medium range, and probabilistic successors such as
GenCast\cite{price2025gencast}, AIFS-CRPS\cite{lang2026aifscrps} and Sonny-ENS\cite{cheon2026sonnyens} produce skillful
ensembles. Run freely for a year, however, many of these models blow up, drift away from the observed climate or lose their seasonal cycle\cite{chattopadhyay2023stability,lehmann2026beyond}. Models built for climate time scales, such as
ACE and ACE2\cite{wattmeyer2023ace,wattmeyer2025ace2}, LUCIE\cite{guan2025lucie}, Spherical
DYffusion\cite{cachay2024dyffusion}, the hybrid NeuralGCM\cite{kochkov2024neuralgcm} and Rescene\cite{cheon2026rescene},
run stably for decades to a century, but they are atmosphere models driven by prescribed sea surface temperature (SST),
the setting that the first intercomparison of AI climate models adopts\cite{henn2026aimip}.

Coupling such atmospheres to the slower components has begun only recently. ACE2-SOM couples ACE2 to a slab ocean and
reproduces equilibrium responses to \co\cite{clark2025acesom}; SamudrACE couples ACE2 to an emulated ocean and sea ice
trained on a physics-based coupled model\cite{duncan2026samudrace,wu2026e3sm}; DLESyM couples deep-learning atmosphere and
ocean models trained on reanalysis and runs for a millennium in the present climate\cite{cresswellclay2025dlesym}; and
ACE2-NEMO and FASST pair learned atmospheres with physics-based ocean models\cite{antonio2026acenemo,aggarwal2026fasst}.
In these systems the El Ni\~no--Southern Oscillation (ENSO), the leading mode of interannual climate
variability\cite{mcphaden2006enso,timmermann2018complexity}, is weaker than observed or is learned from a physics-based
model rather than from observations, and none of them routes the land's water to rivers, runs from emissions through a carbon cycle or follows the climate to the people exposed to it.

Three obstacles stand in the way of an observation-based AI Earth system model. First, an AI atmosphere run freely
drifts, and it cannot represent a climate outside the period it was trained on. Second, the observational record of the
ocean, sea ice and land is too short and incomplete to train their learned counterparts. Third, the coupled variability
is weak: a deterministic surface driven by an atmosphere that is itself held near its climate receives less
high-frequency forcing than the real ocean, and the Bjerknes feedback between equatorial winds and SST that makes ENSO
grow\cite{bjerknes1969} is not enough to sustain El Ni\~no at its observed amplitude\cite{antonio2026acenemo}.
Physics-based modeling met similar problems with anomaly- and flux-corrected coupling\cite{sausen1988flux}, with
intermediate models whose equatorial ocean carries only the wave dynamics of ENSO\cite{zebiak1987model}, with stochastic
climate models in which weather noise integrated by the ocean sustains slow
variability\cite{hasselmann1976stochastic,penland1995optimal}, and with energy-balance models that explain the regional
pattern of forced warming from feedbacks and heat transport\cite{hwang2010increasing,armour2016southern}.

Here we present \model, an Earth system model that keeps a trained AI weather model frozen as its fast component and
builds the slow components around it from such physical formulations, each with few terms and constrained by observations
(Fig.~\ref{fig:model}). It grows out of Rescene\cite{cheon2026rescene}, which made a frozen weather model stable for a
century by blending its large scales toward a fixed calendar climatology at every step. That anchor keeps an AI
atmosphere stable, but it also removes the slow anomalies through which the ocean and land act on the atmosphere, and with a prescribed ocean Rescene has no El Ni\~no, no teleconnections and no forced change. \model\ replaces the step-by-step anchor with a weak
relaxation toward a target that moves with the forced change, takes the SST as an input of its network, and couples it
to an ocean, sea ice, land, rivers, a carbon cycle and an energy balance. We show what each coupling adds, and that the
full model runs for a thousand years without drift, grows and predicts El Ni\~no, reproduces the observed warming, carries a
physically structured forced change around the frozen network, delivers the land's runoff to the rivers and projects
the exposure of people to heat, drought and river floods, at about 1,000 simulated years per day on a single graphics
processing unit (GPU).

\section{Results}

\subsection{\model\ couples a frozen AI atmosphere to physical slow components}

\begin{figure}[!t]
\centering
\includegraphics[width=\textwidth]{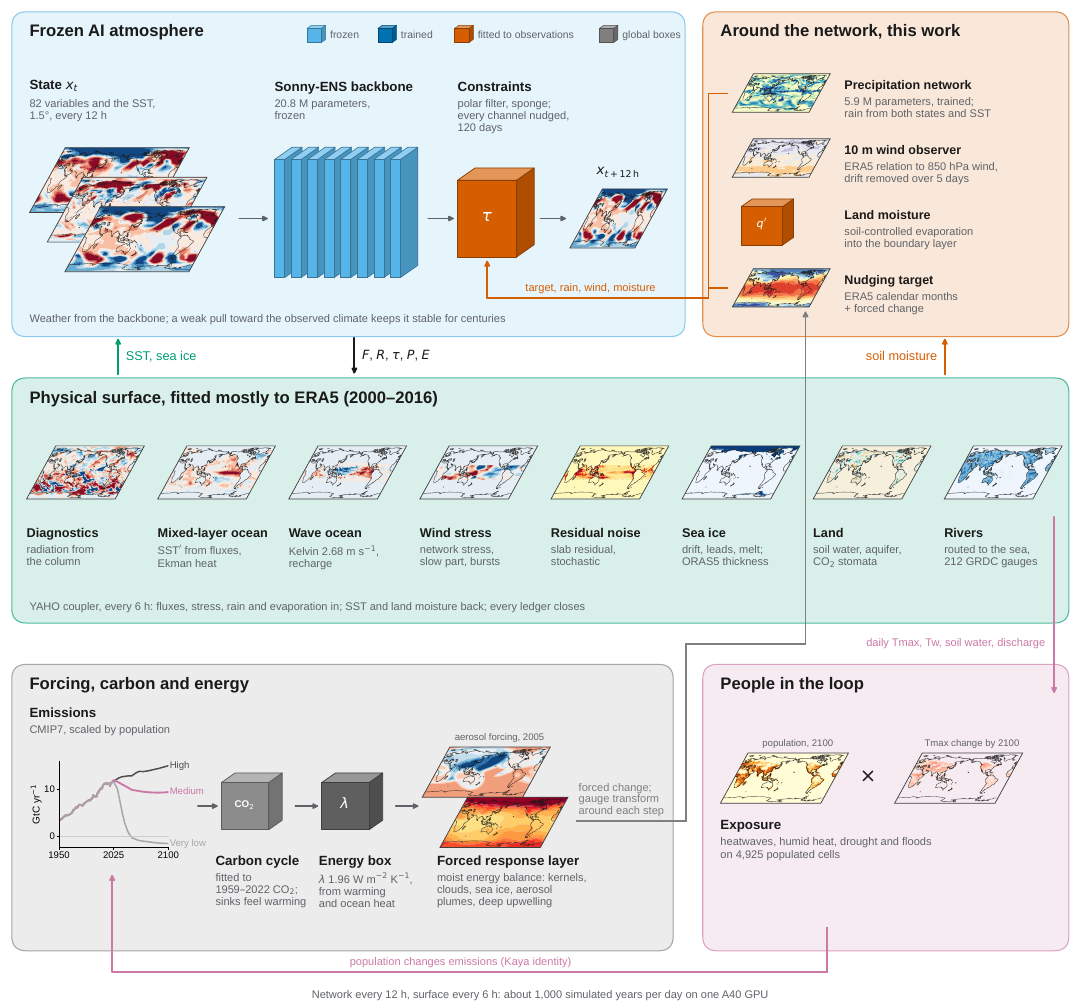}
\caption{\textbf{\model\ and its coupler, YAHO.} The frozen AI atmosphere (blue): Sonny-ENS-S steps 82 variables at 1.5$^\circ$ every 12\,h
from the state and the coupled SST, under a polar filter, a stratospheric sponge and a 120-day relaxation toward the
calendar-month climatology plus the forced change. Around the network (orange): a precipitation network, a low-pass
observer of the 10-m wind and the land's soil-controlled moisture flux. The physical surface (green), fitted mostly to ERA5 over 2000--2016 (Methods): radiation diagnostics, the anomaly mixed-layer ocean with Ekman heat transport, the equatorial wave
ocean, the wind stress for the ocean (the network's weather-scale stress, an SST-forced slow part and westerly wind
bursts), the stochastic residual of the SST equation, sea ice, land and rivers, whose routing has no fitted constant; every 6\,h YAHO passes it fluxes,
stress, rain and evaporation and returns the SST and the land's moisture, with a ledger for each exchange. Forcing, carbon and energy (gray): emissions scaled by
population, a carbon cycle fitted to the 1959--2022 \co\ record, a two-layer energy balance and a forced-response layer
that gives the forced change its pattern, carried around each network step by a gauge transform. People in the loop
(pink): daily heat, humid heat, soil water and discharge at 4,925 populated cells. Thumbnails are fields of the model and
its inputs.}
\label{fig:model}
\end{figure}

The atmosphere is Sonny-ENS-S\cite{cheon2026sonnyens}, a 20.8-million-parameter transformer that steps 82 ERA5 variables
at 1.5$^\circ$ every 12\,h from the state and the SST; it is not retrained or fine-tuned (Fig.~\ref{fig:model}). A polar
filter, a stratospheric sponge and a relaxation toward the calendar-month climatology plus the forced change, with an
e-folding time of 120 days, keep it bounded, and three closures fitted to observations give it what a network that
never saw the soil or a free run lacks: a precipitation network, a low-pass observer for the 10-m wind and the
soil-controlled moisture flux of the land. A coupler, YAHO (Yoking AI atmospheres to Hydrology and Ocean), passes heat,
water and momentum between the network and the surface components every 6\,h, keeps a ledger of every exchange, removes
the mean drift with a flux adjustment and carries the forced change around each network step.
The ocean, an anomaly mixed layer over an equatorial reduced-gravity wave ocean, receives the turbulent and radiative
fluxes and a wind stress that combines the network's weather with an SST-forced slow part and westerly wind bursts, plus
the residual of its own SST equation on ERA5 as a stochastic tendency, and returns the SST. Sea ice drifts, freezes and
melts; the land, a soil-water bucket with groundwater and stomatal closure under \co, returns its moisture flux to the
lowest 150\,hPa of the atmosphere; rivers carry the runoff to the sea, and reservoirs act on the recorded runoff. YAHO's ledgers close: the land's
water to $-0.0016\%$ of its precipitation over a century. Once a model week the carbon cycle and a two-layer energy balance update the size of the forced change, and a moist energy balance, the forced-response layer, gives it a pattern. The relaxation target and the surface climatologies move with it, and a thermodynamic gauge transform carries it around each network step, so the frozen network never sees a climate outside its training. A population recorder keeps the daily maximum
temperature, humid heat, soil water and river discharge at 4,925 populated cells. All components run on the GPU with the
network: a two-member, 100-year control takes 4.9\,h on one NVIDIA A40, about 1,000 simulated years per day across
the two members, two orders of magnitude more than CMIP6-class ESMs simulate on hundreds to thousands of
cores\cite{balaji2017cpmip}. \ST{1} compares \model\ with Rescene component by component, and \ST{2} lists the main experiments.

\subsection{Each coupling restores what the atmosphere alone lacks}

\begin{figure}[!t]
\centering
\includegraphics[width=\textwidth]{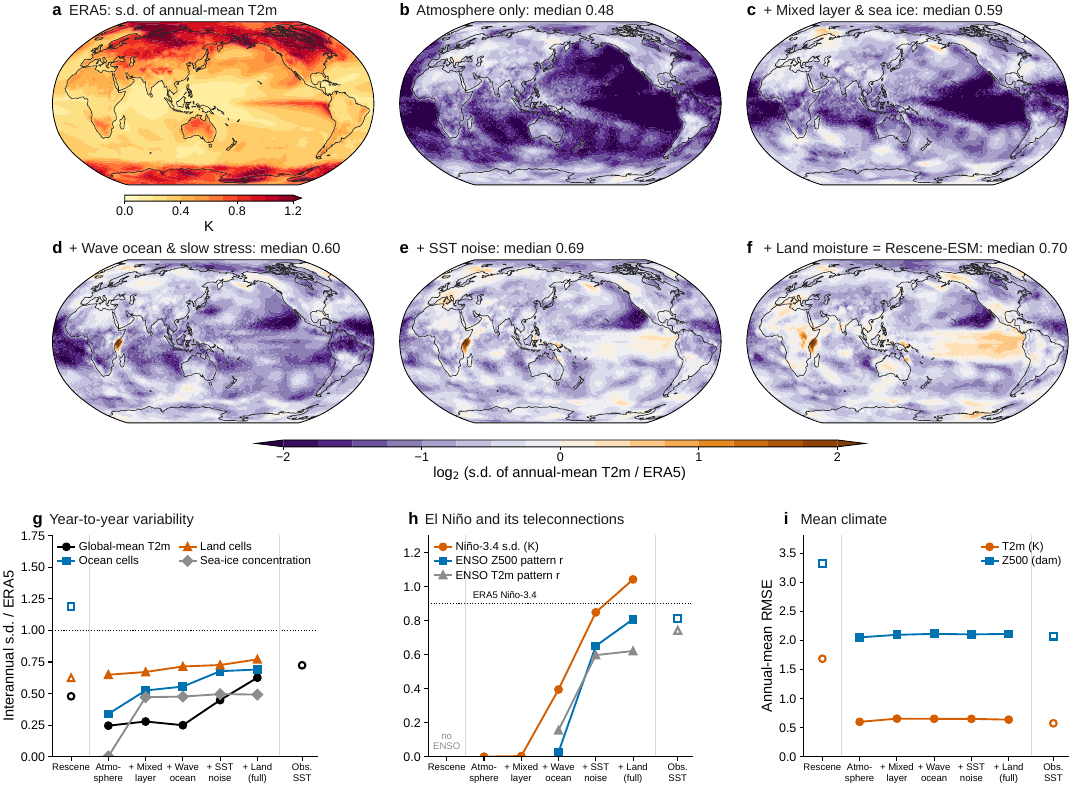}
\caption{\textbf{What each coupling adds.} The couplings are added cumulatively to the atmosphere-only model (SST and
sea ice held at their calendar climatology): the anomaly mixed layer and sea ice, the equatorial wave ocean with the
SST-forced slow stress and wind bursts, the residual of the SST equation as stochastic forcing and the land's moisture flux
(the full model). Each rung is a 30-year, two-member control scored over model years 3--30; the full model uses the same
years of its 100-year control. \textbf{a}, Standard deviation of detrended annual-mean T2m in ERA5 1979--2018.
\textbf{b--f}, $\log_2$ ratio of the same standard deviation in each rung to that of ERA5; titles give the median ratio.
\textbf{g}, Year-to-year variability relative to ERA5: global-mean T2m, the median over ocean and over land cells, and sea-ice concentration. \textbf{h}, Ni\~no-3.4
standard deviation (dotted, ERA5) and the pattern correlation with ERA5 of the December--February Z500 and T2m
regressions on each run's own Ni\~no-3.4. \textbf{i}, Annual-mean T2m and Z500 RMSE. Open symbols are references, not
rungs: Rescene (left; its 100-year run, no ocean and hence no El Ni\~no) and the same atmosphere driven by observed SST
over 1979--2022 (right; AMIP, eight members).}
\label{fig:ladder}
\end{figure}

Adding the couplings one at a time to the same frozen network shows what each contributes (Fig.~\ref{fig:ladder};
\ST{3}). Each rung is a 30-year, two-member control scored over the same years. Over a climatological ocean, the AI
atmosphere keeps the mean climate of its relaxation target (annual-mean error of the 2-m temperature, T2m, 0.60\,K and of the 500-hPa geopotential height, Z500, 20.5\,m) but its year-to-year
variability collapses: annual-mean T2m varies by a median 0.34 of ERA5 over the ocean and 0.65 over land, and the global
mean by 0.25. An interactive mixed layer and sea ice raise the ocean ratio to 0.53 and give sea-ice concentration 0.47 of
its observed variability, but no El Ni\~no. The equatorial wave ocean with the SST-forced slow wind stress produces El
Ni\~no at less than half its observed amplitude (Ni\~no-3.4 standard deviation 0.40\,K) and a teleconnection that these 30 years barely resolve (Z500 pattern correlation 0.03; an earlier configuration of about the same amplitude reaches 0.69 over 100 years, \ST{4}). The residual of the SST equation on ERA5, added as stochastic forcing (\SI{1}), brings
Ni\~no-3.4 to 0.85\,K and the Z500 and T2m teleconnections to 0.65 and 0.60, and raises the ocean ratio to 0.68 and the
global mean to 0.45. The land's moisture flux into the boundary layer raises the land ratio to 0.77, the global mean to
0.63 and the Z500 teleconnection to 0.81, matching the same atmosphere driven by observed SST (0.81), with Ni\~no-3.4 at 1.04\,K in these 30 years (30-year windows of the 1000-year control range over 0.68--1.18\,K). The couplings leave the mean climate almost unchanged: the T2m and Z500 errors stay within 0.60--0.65\,K and 20.5--21.1\,m at every rung, against 1.69\,K and
33.3\,m for Rescene, whose global mean varies by 0.48 of the observed
(Fig.~\ref{fig:ladder}g,i).

\subsection{The coupled climate holds the observed mean state for a thousand years}

\begin{figure}[!t]
\centering
\includegraphics[width=\textwidth]{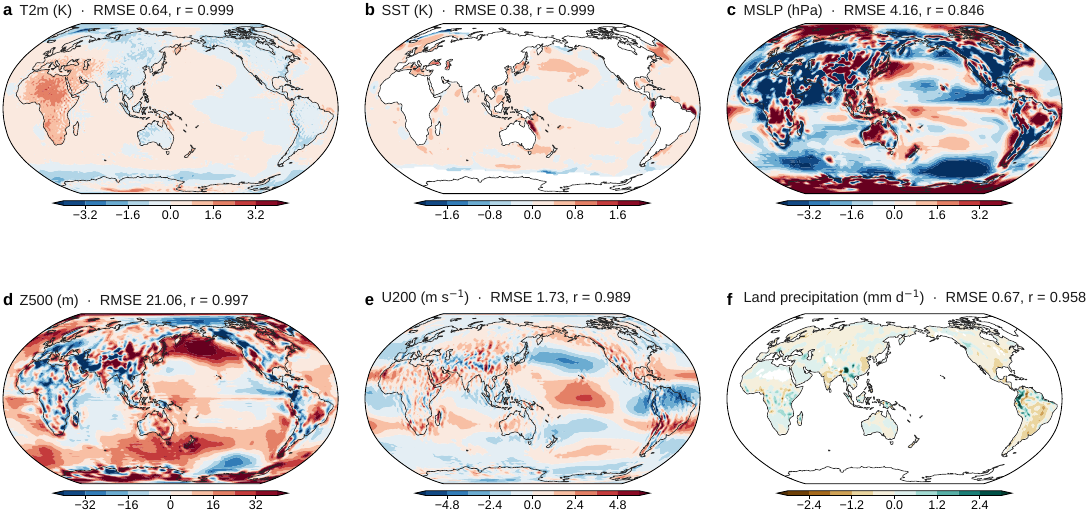}
\caption{\textbf{The coupled mean climate.} Annual-mean departure of the 100-year control (model years 3--100, two
members) from ERA5 1979--2018: \textbf{a}, 2-m temperature; \textbf{b}, SST; \textbf{c}, MSLP; \textbf{d}, 500-hPa
geopotential height; \textbf{e}, 200-hPa zonal wind; \textbf{f}, land precipitation. Titles give the area-weighted RMSE
and the pattern correlation with ERA5.}
\label{fig:climate}
\end{figure}

In a 100-year control without forced change, global-mean T2m drifts by $+0.00$\,K per century,
open-ocean SST by $-0.01$\,K and Arctic and Antarctic sea-ice extent by $+0.00$ and $+0.04\times10^{6}$\,km$^2$ per
century, and the top-of-atmosphere imbalance averages $+0.035$\,\wm\ (\ED{1}; \SI{2}). Against the ERA5 1979--2018
climatology, toward which the relaxation pulls the network's state, the annual-mean T2m error is 0.64\,K (pattern correlation 0.999), SST 0.38\,K, Z500
21.1\,m, 200-hPa zonal wind 1.73\,m\,s$^{-1}$ and land precipitation 0.67\,mm\,day$^{-1}$
(Fig.~\ref{fig:climate}). These errors are within 12\% (T2m) and 2\% (Z500) of those of the same atmosphere driven by
observed SST (0.57\,K and 20.7\,m) and well below those of Rescene (1.69\,K and 33.3\,m). The mean sea-level pressure (MSLP)
error of 4.2\,hPa sits over high terrain and is the same with observed SST (4.25\,hPa; Rescene, with a different network, 0.91\,hPa); it comes from the network's own reduction to sea level. On a Taylor diagram\cite{taylor2001diagram} the time-mean departures from the zonal mean of T2m, SST, 850-hPa
temperature and humidity and the 850-hPa, 200-hPa and 10-m winds correlate with ERA5 at 0.95--0.99 with standard
deviations within 6\% (\ED{2}). Sea ice follows the observed seasonal cycle, from 14.9 to $6.0\times10^{6}$\,km$^2$ in
the Arctic (ERA5, 15.3 and 6.4), the seasonal cycle is reproduced at every latitude (errors of 0.26\,K in zonal-mean SST
and 0.50\,K in T2m) and the annular modes appear as the leading patterns of extratropical MSLP variability (pattern
correlations 0.94 and 0.99) (Supplementary Figs.~3--5; \ST{4}). Sea-ice concentration varies by 0.44 and 0.47 of ERA5 in the Northern and Southern Hemispheres (\SI{11}).

A single-member 1000-year control, which took 15.9\,h on one A40 (about 1,500 years per day with one member and a lighter record), holds the same state over ten times the length
(\SI{20}). Global T2m trends by $-0.0001\pm0.0009$\,K per century (standard error) and its 100-year means stay within
0.04\,K of each other, open-ocean SST trends by $-0.0002\pm0.0006$\,K per century, the annual-mean sea-ice area by
$+0.002$ and $-0.001\times10^{6}$\,km$^2$ per century in the Northern and Southern Hemispheres, and the annual-mean T2m
and SST errors remain 0.64--0.65\,K and 0.37--0.39\,K in every century. Land water closes to $-0.008\%$ of its
precipitation over the 1000 years. The relaxation holds the global mean, so little variance lies at the longest time
scales: 30-year means of global T2m vary by 0.014\,K, and 3\% of the variance of its annual values lies beyond 100 years. A 2-K warm perturbation of the open-ocean SST decays with an e-folding time of 3.1 months, so the coupled system returns to its own climate (\SI{17}).

\subsection{El Ni\~no emerges and is predicted 11 months ahead}

The ocean component alone, driven by prescribed winds as in the experiments of the data-driven ocean model
KIST-Ocean\cite{kim2026kistocean}, carries the dynamics of ENSO (\ED{3}). A 5-day westerly wind burst in the western
Pacific launches a Kelvin wave that crosses the basin at the wave ocean's speed of 2.68\,m\,s$^{-1}$ and warms Ni\~no-3.4 by
0.10\,K a month and a half later; a persistent wind-stress curl pumps the thermocline by 19\,m in 10 days. Driven by ERA5
wind stress and fluxes from the observed state, the ocean reproduces the 2014/15 and 2015/16 El Ni\~no (correlations 0.88
and 0.94) and, outside its fit period, the 1997/98 and 2018/19 events (0.92 and 0.92), all of which fade under
climatological forcing.

Coupled to the atmosphere, this ocean makes its own El Ni\~no. No ENSO statistic enters any fit, though the final configuration was chosen among candidates partly on its ENSO (Methods). Over the 100-year control the Ni\~no-3.4 standard deviation is 0.91\,K
(ERA5 1979--2022, 0.90\,K), the spectrum peaks at 5\,years (5.5) and the variance peaks in boreal winter with a
December-to-May ratio of 2.07 (2.11) (\ED{4}; \SI{6}). Over the 1000-year control the standard deviation is 0.96\,K with
no trend, ranging over 0.84--1.13\,K between centuries and 0.68--1.18\,K (5--95\%) between 30-year windows, so a record of a few decades constrains the amplitude only within that range\cite{wittenberg2009records}; 66\% of the variance lies at periods of 2--7 years, with a peak near 4.4 years (\SI{20}b,c), the variance still peaks in December, and the skewness is $+0.79$ ($+0.51$ to $+1.19$ by century; ERA5, $+0.51$). The SST, wind-stress and thermocline patterns of ENSO match
ERA5 and ORAS5 with pattern correlations of 0.89, 0.94 and 0.83, the Ni\~no-4 stress per kelvin of Ni\~no-3.4 SST is
$1.58\times10^{-2}$\,N\,m$^{-2}$ (1.56; \ED{4}m) and the turbulent heat flux damps Ni\~no-3 SST at 9.7\,\wm\,K$^{-1}$ (11.7)
(\SI{7}). The winter Z500 and T2m teleconnections, regressed on the model's own Ni\~no-3.4 index, correlate with ERA5 at 0.85 and 0.65 (\ED{5}). The Z500 pattern matches ERA5 as closely as that of the same atmosphere driven by observed SST (0.84 over the 100-year control against 0.81 by the measure of Fig.~\ref{fig:ladder}), and the tropical precipitation response has a pattern correlation of 0.79 (\SI{8}). Per 40 years the
model produces 8.2 El Ni\~no and 11.0 La Ni\~na events (ERA5, 10.0 and 10.9) (\SI{9}), and the Pacific tripole index of
decadal variability\cite{henley2015tripole} varies by 0.75\,K (0.78) and, after a 13-year low-pass filter, by 0.13\,K (0.11) (\SI{10}).

Eight-member hindcasts started on 1 May of each year 1997--2021 from ERA5 and a wave-ocean state nudged to ORAS5
thermocline depths before each start predict the Ni\~no-3.4 index with anomaly correlations of 0.77, 0.72 and 0.53 at
leads of 6, 8 and 12 months (persistence, 0.04--0.22 beyond two months), with errors below persistence at every lead and
below climatology through lead 11 (1.04\,K against 1.30\,K at lead 8) and an ensemble spread that grows with the error
(0.79\,K at lead 8) (\ED{6}; \ST{5}). The five starts of 2017--2021, outside every regression of the surface components, reach correlations of 0.79 and 0.94 at leads 8 and 12 (errors of 0.95 and 0.56\,K); the final configuration was chosen among candidates partly on these hindcasts (Methods).

\subsection{Daily weather follows the observed climate}

A 40-year control with a daily record reproduces the statistics on which hazards depend. The distribution of daily
precipitation follows ERA5 up to about 100\,mm\,day$^{-1}$, with 99th percentiles of 22.6 and 39.5\,mm\,day$^{-1}$ over the globe and the tropical ocean (24.4 and 40.8) and a global 99.99th percentile of 85\,mm\,day$^{-1}$ (101), and the Indian summer monsoon has the observed annual cycle (correlation
0.99) and onset (day 150 against day 149) (\ED{7}). Winter blocking in the Northern Hemisphere\cite{davini2012blocking}
has the observed frequency (ratio 0.99) and a pattern correlation of 0.72 with ERA5 (\SI{12}), and western North Pacific
tropical cyclones peak in the observed season and number 17.3 per year below 994\,hPa (ERA5, 15.3) and 7.7 below
985\,hPa (8.1) (\SI{13}; \ST{6}). In the historical runs over 1991--2020, the daily maximum temperature about its 31-day running mean has a median standard deviation of 2.35\,K across the populated cells (ERA5, 2.37), a median lag-1 autocorrelation of 0.63 (0.62) and a median 99th percentile of 5.44\,K (5.34), and heatwave days, each series against its own percentiles, average 6.8 per person in 1986--2005 (6.8).

\subsection{The model follows the warming observed where people live}

\begin{figure}[!t]
\centering
\includegraphics[width=\textwidth]{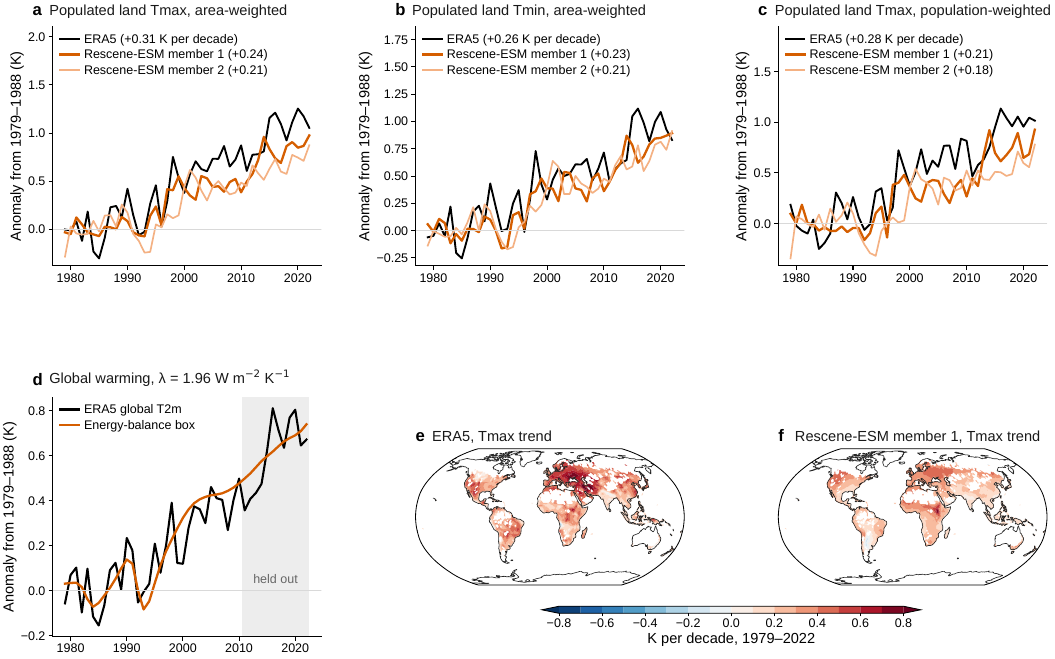}
\caption{\textbf{Historical warming, 1979--2022.} \textbf{a--c}, Annual means over populated land of the daily maximum
(\textbf{a}, \textbf{c}) and minimum (\textbf{b}) temperature, area-weighted (\textbf{a}, \textbf{b}) and
population-weighted (\textbf{c}), as anomalies from 1979--1988, for ERA5 and the two members of the first historical run; legends give linear
trends. \textbf{d}, Global T2m of ERA5 and of the energy balance with $\lambda=1.96$\,\wm\,K$^{-1}$; 2011--2022 (shaded) was withheld from the temperature constraint. \textbf{e}, \textbf{f}, Trend of the annual daily-maximum temperature over populated
land in ERA5 and in the first member.}
\label{fig:hist}
\end{figure}

Three historical runs of two members each cover 1979 to 2022 under observed \co\ and non-\co\ forcing. The energy balance, whose
feedback parameter and aerosol scaling are constrained by the warming of 1979--2010 and the heat uptake of 2006--2020,
gives 1.15\,K above 1850--1900 for 2011--2022, years left out of its temperature constraint, against 1.13\,K in ERA5 (Fig.~\ref{fig:hist}d), and the coupled model's global T2m trend is 0.96 of ERA5's (\ST{1}). Over populated land the annual daily-maximum temperature tracks ERA5 with correlations of 0.92 (area-weighted) and 0.88 (population-weighted) and warms by 0.21--0.24\,K per decade against 0.31\,K, with a trend pattern correlation of 0.36 (Fig.~\ref{fig:hist}e,f), and the daily-minimum temperature warms by 0.21--0.23\,K per decade against 0.26\,K (Fig.~\ref{fig:hist}a--c). Ice-free SST warms by 0.11--0.12\,K per decade (ERA5, 0.085) (\SI{14}). Dry soil comes with hot days as observed: the summer
correlation between monthly soil-water and daily-maximum temperature anomalies has a pattern correlation of 0.62 with ERA5
over land (\SI{15}).

\subsection{Forced change enters with a physically based structure}

\begin{figure}[!t]
\centering
\includegraphics[width=\textwidth]{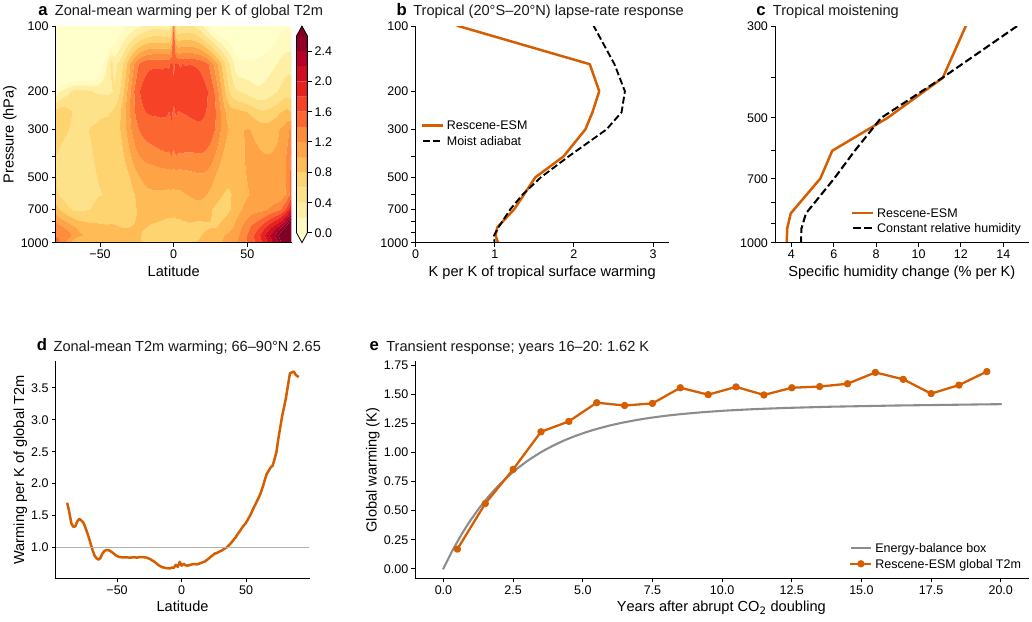}
\caption{\textbf{Structure of the \co\ response.} Abrupt doubling of \co\ against a control at fixed \co, last five of 20
years. The vertical structure and the humidity change are set by the forced-response layer (moist adiabat in the
tropics, fixed relative humidity) and carried around the network by the gauge transform; panels \textbf{b} and
\textbf{c} show that the coupled model keeps them. \textbf{a}, Latitude--pressure section of zonal-mean warming per kelvin of global T2m warming. \textbf{b}, Tropical (20$^\circ$\,S--20$^\circ$\,N)
warming profile per kelvin of tropical surface warming, with the moist adiabat. \textbf{c}, Tropical specific-humidity change per kelvin of global T2m warming, with the change at constant relative humidity. \textbf{d}, Zonal-mean T2m warming per kelvin of global
warming. \textbf{e}, Annual global warming of the model and of the energy balance.}
\label{fig:co2}
\end{figure}

The frozen network never steps a climate it was not trained on. In forced runs the energy balance sets the magnitude of
the warming, the forced-response layer gives it a pattern and a vertical structure (a moist adiabat in the tropics and
fixed relative humidity), and the gauge transform carries this change around each network step, so the network steps the
weather of the present climate while the state carries the forced change (Methods). After an abrupt doubling of \co\
(Fig.~\ref{fig:co2}) the tropical troposphere therefore warms 2.14 times as much at 300\,hPa as at the surface (moist
adiabat, 2.41) and the tropical 850-hPa specific humidity rises by 4.0\% per kelvin of global warming (at fixed relative humidity, 4.7\%), and the coupled model holds this structure for 20 years without relaxing back toward its training
climate. Land warms 1.30 times as much as the ocean and the Arctic north of 60$^\circ$\,N 2.40 times the global mean
(\ST{7}), from the moist energy
balance and the sea-ice albedo of the forced-response layer, none of whose constants is fitted to warming trends. Global
precipitation, which the precipitation network diagnoses from the warmer and moister states, rises by 1.52\% per
kelvin\cite{held2006robust,allen2002constraints} (\ED{8}). The warming patterns of 2$\times$,
3$\times$ and 4$\times$\co\ (1.60, 2.51 and 3.13\,K; \ST{7}) correlate at 0.98, and the transient climate response under a 1\%-per-year increase is 1.60\,K (\SI{16}).

\subsection{Land water reaches the rivers}

The runoff of the historical run, routed along the DDM30 network\cite{doll2002ddm30}, reaches 212 gauges of the Global Runoff Data Centre (GRDC) with a
median ratio of simulated to observed mean flow of 1.15 and a median seasonal-cycle correlation of 0.79 (0.90 for the
Amazon, 0.83 for the Mississippi and 0.75 for the Paran\'a); no constant is fitted to discharge (\ED{9}). Routed through
the 7,302 reservoirs of GRanD and ISIMIP3a\cite{lehner2011grand,frieler2024isimip3a} with the operating rule of Hanasaki
et al.\cite{hanasaki2006reservoir}, the median seasonal range of the 129 regulated gauges falls from 1.84 to 1.06 times
the mean flow, against 1.03 observed (\ED{10}).

\subsection{Emissions and population set future exposure}

\begin{figure}[!t]
\centering
\includegraphics[width=0.86\textwidth]{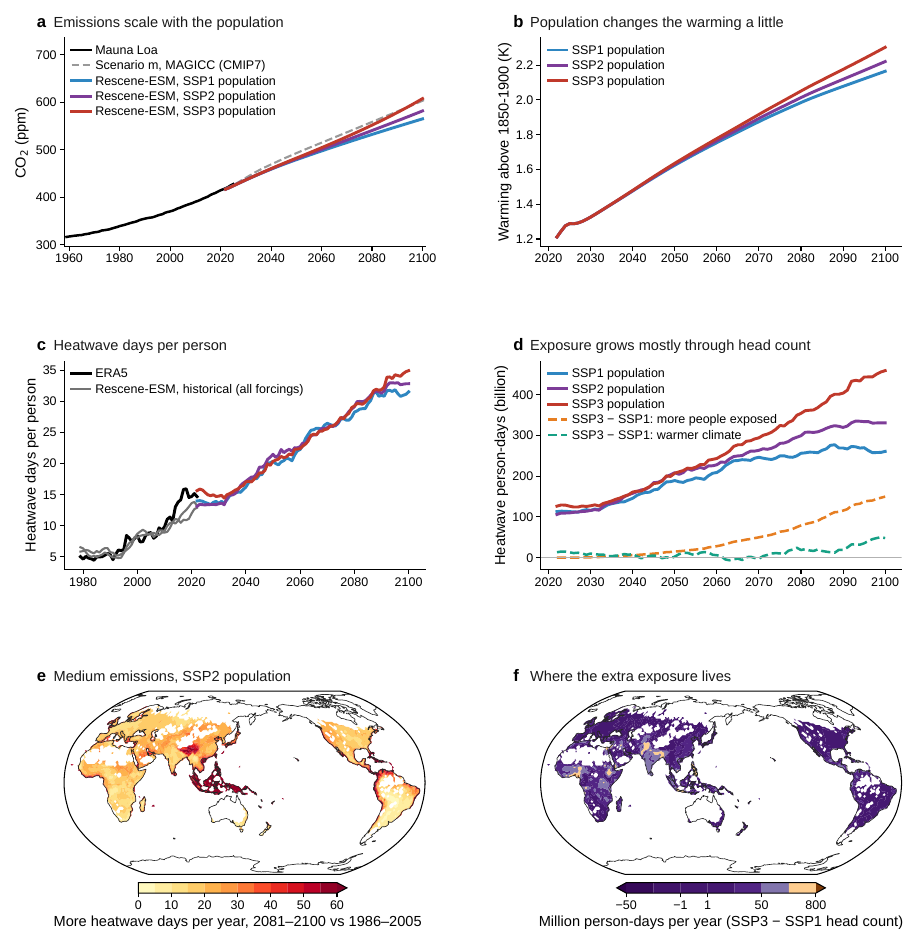}
\caption{\textbf{Emission-driven futures and the people exposed.} The CMIP7 medium emissions scenario with emissions
scaled by the SSP1, SSP2 and SSP3 population pathways. \textbf{a}, Atmospheric \co\ from the carbon cycle, with Mauna Loa
and the scenario's MAGICC concentration. \textbf{b}, Warming above 1850--1900. \textbf{c}, Heatwave days per person, ERA5
and the historical runs, then the three pathways. \textbf{d}, Heatwave exposure and the split of the SSP3--SSP1 difference
into the part from the number of people and the part from the warmer climate. \textbf{e}, Change in heatwave days per
year, 2081--2100 against 1986--2005, under SSP2. \textbf{f}, Where the additional exposure of SSP3 against SSP1 lives.}
\label{fig:people}
\end{figure}

Driven by the CMIP7 medium emissions scenario through its carbon cycle, \model\ continues from 2022 with atmospheric
\co\ of 416.6\,ppm (Mauna Loa, 418.5\,ppm) and reaches 581\,ppm in 2100, 3.5\% below the scenario's MAGICC
concentration\cite{meinshausen2011magicc} (Fig.~\ref{fig:people}a). Emissions scale with population through per-capita
emissions, so the SSP1 and SSP3 population pathways\cite{kc2017ssp} (8.2 and 13.2 billion people in 2100) reach 565 and
607\,ppm and warm by 2.08 and 2.18\,K above 1850--1900 in 2081--2100 (SSP2, 2.12\,K) (Fig.~\ref{fig:people}b). Heatwave
days, counted as in the Lancet Countdown\cite{romanello2023lancet}, rise with ERA5 over 1979--2022 and exceed 30 per
person per year by 2100 under all three pathways (Fig.~\ref{fig:people}c). Exposure reaches 265, 323 and 410 billion person-days per year in 2081--2100; of the 145 billion between SSP3 and SSP1, 118 billion come from the number of people exposed and 27 billion from the warmer climate their emissions produce (Fig.~\ref{fig:people}d--f).

With the very low and high scenarios' own emissions, non-\co\ forcing and aerosol patterns, \co\ peaks at 441\,ppm
in 2038 and falls to 397\,ppm in 2100 under very low emissions and reaches 668\,ppm under high emissions (MAGICC, 401
and 697\,ppm), and the warming in 2081--2100 is 1.14 and 2.43\,K above 1850--1900 (\ED{11}a,b). Paired with their own
population pathways, the SSP1 and SSP3 worlds reach 104 and 487 billion heatwave person-days per year in 2081--2100, 12.0 and 38.3 days per person (\ED{11}c); of the difference of 383 billion, 290 billion come from the warmer climate and 93 billion from the number of people (\ED{11}d--f). Between consistent worlds, emissions set three-quarters of the difference in heat exposure; within the medium emissions scenario, the number of people sets four-fifths of a difference less than half as large.

Water changes as well. Under the medium pathway, against 1995--2014, mean river flow rises at 76\% of the 241 routed gauge sites (median $+14\%$) and the annual maximum daily flow at 63\% (median $+12\%$), river-flood days per person rise from 1.78 to 2.28 and drought months per person from 1.19 to 1.26 (\SI{18}; \ST{7}).

\section{Discussion}

\model\ shows that an AI weather model trained only for medium-range forecasting can serve, frozen, as the atmosphere of
an Earth system model. Its climate is stable for a thousand years, its El Ni\~no is of the observed size and rhythm and is predictable 11 months ahead, its daily weather carries the observed monsoon, blocking and cyclones, and forced change
reaches it with a physically based structure without retraining. Three ideas make this work. A weak relaxation toward a target that moves
with the forced change keeps the frozen network bounded while letting the slow anomalies of the ocean and land act on it,
and a gauge transform carries the forced change around the network, so no retraining is needed for a warmer world. Each slow component is a physical equation with few terms, constrained by observations, and the part of the tropical SST equation that observations do not explain returns as stochastic forcing. And YAHO keeps a ledger of every exchange, so components can be added one at a time, as in the coupling ladder, with the budgets kept.

Where they overlap, the exposure of \model\ agrees with multi-model projections, and it extends them past mid-century.
Over 2023--2050 a warm-spell index counted as in a study of 19 CMIP6 models\cite{dash2025compound} rises by a factor of 2.2--2.4 from 1991--2020 (2.0--2.1 weighted by population), within the factor of 2--3 reported there for 2021--2050 (\ST{9}). The rise in heatwave exposure comes mainly from the climate (62--64\%, with 19\% from population and 17--19\% from their product), almost wholly in Europe (97--98\%) but only a third in Africa, and in Asia and Africa it is largest in the SSP3 world, as for the heat-related compound extremes of that study. The three worlds then separate: within 8\% of each other over 2023--2050 (139--150 billion person-days per year), they differ 4.7-fold in 2081--2100 (104--487 billion). Under high emissions the product of more heat and more people is by then the largest part of the rise from 1986--2005, 49\% worldwide and 75\% in Africa. Heatwave days measure departures from each place's own climate, and physiological limits place the exposure differently. Age-specific wet-bulb limits raise the number exposed two- to eightfold and, at 3\,K of warming, place 73\% of those exposed for at least 180 hours a year in India, China, Pakistan and Bangladesh\cite{kong2026heat}. These four countries hold 23--25\% of the heatwave person-days here under medium and high emissions (2.1--2.4\,K) and 29\% of the people. On the
1.5$^\circ$ grid ERA5 itself exceeds the young-adult limit of 30.6$^\circ$C on at most 0.3 days per year in any populated
cell, so absolute limits and age-structured population call for a finer exposure grid.

The speed of about 1,000 simulated years per day on one GPU makes the experiments that impact studies need routine:
large initial-condition ensembles, many emissions and population pathways, and their consequences for rivers and
people, all within the same model. The measured departures from the observations locate the next gains. In the atmosphere they are the MSLP error of 4.2\,hPa over high terrain inherited from the network, 2.9 storms deeper than 975\,hPa per year against 4.1 and a hydrological sensitivity that falls from 1.52\% to 0.85\% per kelvin between 2$\times$ and 4$\times$\co. In the ocean and ice they are an ENSO skewness of $+0.79$ over 1000 years against $+0.51$, sea-ice variability at 0.44--0.47 of ERA5, 30-year means of global T2m that vary by only 0.014\,K and the SST variability of the western boundary currents at 0.4 of ERA5, which the extratropical residual of the SST equation raises to 0.6--0.7 at 0.7 of its fitted size, with Ni\~no-3.4 at 0.86\,K over 30 years and a trend of $+0.20$\,K per century (\ST{8}). In the forced response they are populated-land daily-maximum temperature trends of 0.21--0.24 against 0.31\,K per decade and an equilibrium warming of 1.9\,K for doubled \co\ from the feedback fitted to the observed warming. Coupling a three-dimensional ocean to the same frozen atmosphere is the next step.

\section{Methods}

\subsection{Model overview and the YAHO coupler}
The coupler, YAHO (Yoking AI atmospheres to Hydrology and Ocean), connects the frozen network to the surface and
forced-change components; it reads and writes the network's state by variable name, so the same coupler also runs the
Rescene atmosphere\cite{cheon2026rescene}. \model\ runs its surface on the 1.5$^\circ$ latitude--longitude grid (120\,$\times$\,240 cells, rows from
88.5$^\circ$\,S) with a 6-h coupling step; the network's grid has one more row at the South Pole, which the coupled state
drops. In every second 6-h step the network takes one 12-h step from the SST that the ocean holds at its start; the 6-h state in between is the mean of the two 12-h states and is used only for the exchange with the surface
(diagnostics of extremes use the network's own 12-hourly states). In each 6-h step (1) the surface fluxes are computed
from the atmospheric state with bulk formulas\cite{large2004forcing} and learned radiation diagnostics, (2) the ocean,
sea ice, land and rivers advance and the flux adjustment is subtracted, (3) the land's moisture flux enters the lowest
150\,hPa of the atmosphere and (4) the ledgers of heat and water are updated. The forced-change components update
once a model week. Every component is written in PyTorch and runs on the GPU with the network, which runs under
half-precision autocast. Members differ by the network's dropout draws and by the seeds of the stochastic terms.

\subsection{Atmosphere}
The atmosphere is Sonny-ENS-S\cite{cheon2026sonnyens}, the 20.8-million-parameter variant of Sonny-ENS, the ensemble
successor of Sonny\cite{cheon2026sonny}. It predicts the 12-h tendency of 82 variables on the 1.5$^\circ$ WeatherBench~2
grid\cite{rasp2024wb2} (geopotential, temperature, specific humidity, zonal and meridional wind and vertical velocity on
13 pressure levels, T2m, the 10-m wind components and MSLP) from the state, the SST, the land--sea mask, the surface
geopotential and the insolation, with a two-step transformer that resolves the dynamics before the thermodynamics. It
was trained on ERA5 1979--2018 with a deterministic loss and then with the fair continuous ranked probability score
under Monte Carlo dropout, which stays on in \model\ so that members differ. Its weights are not changed.

Run freely, the network keeps its SST response (December--February 1997/98 Z500 pattern correlation 0.83 north of
20$^\circ$\,N in a 16-member prescribed-SST run) but accumulates grid-scale noise in the stratosphere and blows up after
about six months. A Fourier filter poleward of 45$^\circ$, a sponge that relaxes the 50-, 100- and 150-hPa levels toward
the calendar-month climatology over 10, 20 and 40 days, and a relaxation of every channel toward a target with an
e-folding time of 120 days keep it bounded. The target is the ERA5 calendar-month climatology plus the forced change of
the run. With these constraints the prescribed-SST run stays bounded past four years and keeps the Z500 response to the
1997/98 SST at a pattern correlation of 0.76; without the relaxation the coupled model diverges.

The network carries no precipitation. A diagnostic network of 5.9 million parameters gives the 12-h precipitation from
the two states that bound each step and the SST; it was trained on ERA5 1979--2016 with the squared error of
$\log(1+P)$ plus a squared error of the amount and validated on 2017--2019. Under these constraints alone, the network's 10-m wind drifts away from its own 850-hPa wind (zonal-mean meridional wind at 60$^\circ$\,S, $+3.8$\,m\,s$^{-1}$ in 30 years) while the
850-hPa wind stays within 0.2\,m\,s$^{-1}$ of ERA5. Per cell and calendar month, $(u_{10},v_{10})$ is fitted as a linear
function of $(u_{850},v_{850},1)$ on ERA5 2000--2016 (independent skill on 2017--2022, 0.80--0.82). Every 12-h step moves
the 10-m wind toward this relation by $1-e^{-12\,\mathrm{h}/2\,\mathrm{days}}$ of the running mean of the departure
(e-folding 5 days) and removes each move from that running mean, so the slow drift is removed (87\% at 60$^\circ$\,S) while the
weather-scale departure stays the network's. The network never sees the soil, so the land acts on the air as moisture:
the anomaly of the land's evaporation from its 30-day mean, together with its slow departure from the model's own
calendar-month climatology, moistens the lowest 150\,hPa (capped at 60\,\wm\ of latent heat). Between 20$^\circ$\,S and
20$^\circ$\,N (tapered to zero at 35$^\circ$) the bulk fluxes into the ocean read the near-surface air anomalies moved
halfway toward their equilibrium with the SST, the adjustment that the boundary layer makes within a day; the network's
state is not changed.

\subsection{Ocean}
SST is carried as an anomaly on the 2000--2016 calendar climatology,
\begin{equation}
\frac{d\,\mathrm{SST}'}{dt}=a_F F'+a_E Q'_{\mathrm{Ek}}+a_R R'+a_h h'+\mathcal{A}'-\kappa\,\mathrm{SST}'+\sigma\,\xi(t),
\end{equation}
with $F'$ the turbulent heat-flux anomaly against the flux of the climatological air--sea state, $Q'_{\mathrm{Ek}}$ the
Ekman heat advection\cite{codron2012ekman}, $R'$ the net surface radiation anomaly from a diagnostic trained on ERA5, $h'$
the thermocline depth anomaly of the wave ocean and $\mathcal{A}'$ the advection of the climatological SST by the anomalous currents (the wave ocean's and those of the frictional surface layer) and of the anomaly by the climatological surface currents. The coefficients are fitted per cell and calendar month on
ERA5 2000--2016 by instrumental-variable regression, which removes the bias of a same-time regression in which the
atmosphere drives both the fluxes and the SST; $h'$ and the currents come from the wave ocean driven by ERA5 winds, so no
subsurface data enter the fit. Zero anomaly under a climatological atmosphere is an exact equilibrium, and the heat the
ocean receives leaves the atmosphere. A daily second-order Shapiro filter removes the two-cell mode of centered advection.

The last term is the residual of the same equation on ERA5: over non-overlapping 30-day windows of 2000--2016 the
residual is projected on 20 tropical empirical orthogonal functions, and each mode is an Ornstein--Uhlenbeck process
with its own calendar-month standard deviation and decorrelation time. In Ni\~no-3 the residual is 0.40\,K per 30 days (\SI{1}a), nearly white and largest in February--April; it stands in for the weather forcing\cite{hasselmann1976stochastic} and the unresolved ocean processes that a deterministic mixed layer omits.

The thermocline dynamics that a mixed layer cannot carry are represented by a linear 1.5-layer reduced-gravity ocean between
30$^\circ$\,S and 30$^\circ$\,N\cite{mccreary1981linear,zebiak1987model},
\begin{equation}
\partial_t u-fv=-g'\partial_x h+\frac{\tau'_x}{\rho H}-ru,\quad \partial_t v+fu=-g'\partial_y h+\frac{\tau'_y}{\rho H}-rv,
\quad \partial_t h+H\nabla\cdot\mathbf{u}=-rh,
\end{equation}
with a Kelvin-wave speed of 2.68\,m\,s$^{-1}$, $H=150$\,m and $r=(2.5\,\mathrm{yr})^{-1}$, on an Arakawa C grid with closed
coasts and a sponge poleward of 20$^\circ$. Hindcasts start from a wave-ocean reanalysis in which the model is driven by
ERA5 winds and nudged to ORAS5\cite{zuo2019oras5} thermocline-depth anomalies (time scale 30 days) with data before each
start only.

The ocean's stress anomaly is split by time scale. Its weather part is the network's own stress minus its 30-day running
mean. Its slow part is a ridge regression, fitted on ERA5 1979--2016, from the tropical convective heating anomaly that
the SST anomaly of the past 30 days sets through the monotone ERA5 precipitation--SST curve,
$P'=f(\mathrm{SST}_c+\mathrm{SST}')-f(\mathrm{SST}_c)$, to the stress of the next 30 days (scaled by 0.9); the lag removes
the bias of a same-time regression in which the wind drives the SST, and the convective threshold makes the response
asymmetric between warm and cold events, as in ERA5. Westerly wind bursts\cite{lengaigne2004wwb} are added as events
resampled from those detected in ERA5 1979--2022, at the observed rate for the current Ni\~no-3.4 state, with their
long-term mean stress removed.

The experiments of \ED{3} run the ocean component alone, as KIST-Ocean was tested\cite{kim2026kistocean}: the
mixed layer and the wave ocean with the coefficients of the full model and its daily Shapiro filter, forced by
prescribed wind stress. The burst is a sine-by-sine stress patch over 130--170$^\circ$\,E, 10$^\circ$\,S--10$^\circ$\,N
held for 5 days; the curl is a Gaussian stress vortex one grid cell wide at 15.5$^\circ$\,N, 135.5$^\circ$\,E held for 30
days; both are differences from an unforced run. The cases start from the ERA5 SST anomaly and the wave-ocean state
driven by ERA5 winds on the start date, and are forced either by ERA5 stress and surface fluxes or by none.

\subsection{Sea ice}
Sea-ice concentration is an anomaly on the 2000--2016 climatology, driven by the flux anomalies and damped, with both
coefficients fitted per cell on ERA5 2000--2016. Its thermodynamics partition energy with the mixed layer: open water
cooled below freezing forms ice, ice carried by the wind into water that was open melts with the mixed layer's heat above
freezing, leads opened in the pack close with new ice of 0.5\,m\cite{hibler1979dynamic} as air below freezing removes
their heat, and ice beyond the climatological cover melts with the ocean heat flux from below (30\,\wm\ in the Southern
and 2\,\wm\ in the Northern Hemisphere). The ice drifts freely with the wind\cite{thorndike1982sea}, at 2\% of the 10-m wind anomaly turned
20$^\circ$ to the right in the Northern and to the left in the Southern Hemisphere, in flux form with upwind fluxes
and no flux into land. Ice-covered cells hold the SST at freezing.

\subsection{Land, rivers and reservoirs}
The land is a bucket\cite{manabe1969bucket} for root-zone soil water with a degree-day snow store, gravity drainage into a
groundwater store that returns baseflow, a soil heat store whose capacity depends on soil water, and evaporation limited
by the available water. Stomata close with \co\ with the elasticity measured at FLUXNET
sites\cite{pastorello2020fluxnet}, which follows the transpiration share of evaporation in GLEAM\cite{martens2017gleam} (\SI{19}a--c). Runoff is routed on the DDM30 drainage network\cite{doll2002ddm30} upscaled to the
model grid, each cell a linear reservoir with a flow speed of 0.5\,m\,s$^{-1}$ and four substeps a day; river mouths hand the freshwater to the ocean's ledger (the ocean carries no salinity) and inland sinks keep theirs. Rivers do not feed back on the climate, so reservoirs act on the run's recorded daily runoff. Runs record the 241 GRDC gauges matched to the network; the 212 with at least three years of record from 1990 on are evaluated. The ISIMIP3a dams\cite{frieler2024isimip3a} (GRanD v1.3\cite{lehner2011grand} and
additional dams, 7,302 with a capacity, 7.0 thousand km$^3$ in total, on 1,750 river cells) operate from their
completion year with the non-irrigation release rule of Hanasaki et al.\cite{hanasaki2006reservoir}: the provisional
release is the cell's mean annual inflow, scaled each year by the storage relative to 85\% of capacity, and where the
capacity is less than half a year's inflow the release mixes toward the inflow; storage beyond capacity spills. No
constant is fitted to discharge, and the routing without dams reproduces the coupled run's own gauge series (median
relative difference of mean flow, $10^{-5}$). In a sensitivity run (\ED{10}) domestic and industrial consumption from
ISIMIP3a and irrigation, the irrigated area in season (crop calendars\cite{jagermeyr2021crop}) times
$\max(E_0-0.8P,0)$ with FAO-56 reference evapotranspiration\cite{allen1998fao56}, are drawn from reservoirs, then rivers, then
groundwater (1,133\,km$^3$ per year in total, 39\% from groundwater).

\subsection{Flux adjustment}
Each surface component is fitted against the ERA5 atmosphere; coupled to the AI atmosphere it receives slightly different
mean fluxes and, uncorrected, drifts toward a climate of its own. As in flux-corrected coupled
models\cite{sausen1988flux}, the mean drift is removed in two parts. First, coupled runs from initial dates in 2000--2016
record the mean anomaly of every exchanged flux (turbulent heat flux into the ocean, surface and top-of-atmosphere
radiation, land evaporation, and the wind stress within 30 days) by season, lead (0--1, 1--3, 3--7, 7--30, 30--90 and
more than 90 days) and cell; against the 2000--2016 climatologies any mean anomaly is drift. Second, slower drifts are read
from the model's own 30-year control: the calendar-month SST bias against the 2000--2016 climatology, box-smoothed over
five cells, times a heat-flux feedback of 15\,\wm\,K$^{-1}$ in the tropics and 6\,\wm\,K$^{-1}$ at 35--75$^\circ$, is
added to the turbulent heat-flux adjustment from 30 days on, and the top-of-atmosphere ledger receives the matching step.
The adjustment is a fixed function of season, lead and cell: it carries no information on the anomaly of any year, and
forced and interannual changes pass through it unchanged.

\subsection{Forced change}
A two-layer energy balance\cite{held2010probing,geoffroy2013transient} gives the global surface temperature under the
effective radiative forcing of \co\ ($5.35\ln(C/C_0)$\,\wm) and of the non-\co\ agents (Indicators of Global Climate
Change history\cite{forster2024igcc}, CMIP7 scenarios afterward). The feedback parameter and the scaling of the aerosol
forcing are constrained jointly by the observed warming over 1979--2010 and the Earth's heat uptake over 2006--2020
($\lambda=1.96$\,\wm\,K$^{-1}$, aerosol forcing $\times0.65$; equilibrium warming for doubled \co, 1.9\,K); the surface warming of 2011--2022 is left out of the constraint, while the heat-uptake period overlaps it. The 1979--2018 climatology is taken as 0.75\,K above 1850--1900. In emission-driven runs a four-pool carbon cycle whose sinks weaken with cumulative uptake
and warming\cite{smith2018fair}, spun up from 1750, turns emissions into concentration; its pre-industrial 100-year
integrated impulse response ($r_0=27.5$ years) is fitted to the Mauna Loa record\cite{keeling1976maunaloa} of 1959--2022
(error 1.19\,ppm; \SI{19}d).

The regional pattern of forced change comes from a moist energy balance on the sphere,
\begin{equation}
C(\mathbf{x})\,\partial_t T=F(t)+F_{\mathrm{aer}}(\mathbf{x},t)+R(\mathbf{x};T)+\frac{p_s}{g}\nabla\cdot(D\nabla h_m')
-\gamma(\mathbf{x})(T-T_{\mathrm{up}})+O_{\mathrm{Ek}}(T),
\end{equation}
where $h_m'$ is the near-surface moist static energy anomaly at fixed relative humidity,
$D=1.06\times10^{6}$\,m$^2$\,s$^{-1}$\cite{hwang2010increasing}, $R$ the top-of-atmosphere response built from radiative
kernels for temperature, water vapor and surface albedo and from cloud-controlling factors fitted to ERA5,
$F_{\mathrm{aer}}$ the regional departure of the aerosol forcing, $\gamma$ the exchange with the deep ocean where
climatological Ekman pumping upwells water that has not seen the warming, and $O_{\mathrm{Ek}}$ the advection of the mixed
layer's anomaly by the climatological Ekman transport\cite{codron2012ekman,armour2016southern}. Its global mean is nudged
to the energy balance, which keeps the magnitude, and it gives the shape: the relaxation target, the surface climatologies and the sea-ice climatology (in the Arctic melt season through the thinning of the pack, from ORAS5 thickness) move with it every model week, the free troposphere following a moist adiabat
in the tropics and the latitude's vertical profile elsewhere, humidity at fixed relative humidity and geopotential by the
hypsometric equation. The network was trained on the climate of 1979--2018, so in forced runs the forced change is subtracted from the state, the SST and the relaxation target before each network step and added back after it (a
thermodynamic gauge transform): the network steps the weather of the climate it knows while the thermodynamic state
carries the forced change. Emission-driven runs take the CMIP7 ScenarioMIP medium (``m'') \co\ emissions after 2022; for
the SSP1 and SSP3 population pathways the emissions are scaled by the ratio of each pathway's population to that of SSP2,
so population changes both exposure and warming. Two further runs take the very low (``vl'') and high (``h'') scenarios'
\co\ emissions and non-\co\ forcing, with each scenario's regional aerosol pattern from the same plume model; their
aerosol forcing is divided between aerosol--radiation and aerosol--cloud interactions in the proportions of the medium
scenario. For the exposure each is paired with its own population pathway (very low with SSP1, high with SSP3), and the
difference between the two worlds is split as below.

\subsection{Exposure}
Population from the CMIP7 population density (history and SSP pathways at 0.25$^\circ$) is summed onto the model grid
(4,925 populated cells). During a run the recorder keeps the daily maximum and minimum T2m, the daily maximum wet-bulb temperature (a proxy from temperature, pressure and humidity, applied alike to ERA5), the root-zone water and the river discharge at these cells. As for ERA5, the daily extremes come from four values a day: the network's states at 00 and 12\,UTC and, at 06 and 18\,UTC, a diagnostic fitted per cell on ERA5 2000--2016 that adds a diurnal departure to the mean of the two neighboring states. Heatwave days are days in runs of at least two days with both the daily maximum and minimum above their local 1986--2005 95th percentiles\cite{romanello2023lancet}, the model's from the two members of its first historical run; drought months have root-zone water below the calendar month's 10th percentile and flood days have discharge above its 99.5th percentile, over the same base period. The difference in exposure between two pathways is
split exactly into a climate part, the difference in hazard times the mean of the two populations, and a head-count part,
the mean hazard times the difference in population. The rise in exposure from a past period is split as in Dash et
al.\cite{dash2025compound}: climate is the change in hazard times the past population, population is the past hazard
times the change in population, and the rest is their product. The warm-spell index counts days in runs of at least six
days with the daily maximum above its 95th percentile over all days of 1991--2020, with thresholds from the two members of
the first historical run, and is averaged over the populated cells. Countries and continents are those of the nearest
Natural Earth polygon.

\subsection{Coupling ladder}
Four additional 30-year, two-member controls build the model up one coupling at a time, with the same network
constraints, precipitation network, 10-m wind observer and flux adjustment as the full model. The atmosphere-only rung
holds SST and sea ice at their calendar climatology. The mixed-layer rung adds the anomaly mixed layer and sea ice with the full model's coefficients but without the wave ocean's thermocline and current terms. The wave-ocean rung adds the
equatorial wave ocean, the SST-forced slow stress and the westerly wind bursts; the noise rung adds the stochastic SST
residual; and the full model adds the land's moisture flux into the boundary layer, for which the first 30 years of the
100-year control are used. Rescene, the predecessor without interactive components\cite{cheon2026rescene}, is shown from
its 100-year run. All are compared with ERA5 after discarding the first two years.

\subsection{Experiments}
Controls start from the ERA5 state of 1 January 1979 and a spun-up land, without forced change (100 years with two
members; 1000 years with one member, recording annual-mean fields and monthly Ni\~no-3.4; 40 years with a daily record). Historical runs cover 1979--2022 with observed \co\ and non-\co\ forcing (three
runs of two members each). The \co\ experiments run abrupt 2$\times$, 3$\times$ and 4$\times$\co\ for 20 years and a 1\%-per-year increase
for 140 years, each against the time mean of a 20-year control at fixed \co, and a perturbation run adds 2\,K to the open-ocean SST of the
control's initial state for five years. Scenario runs start on 1 January 2022 from the ERA5 atmosphere and ocean and the
land and river state that a historical run from 1979 reaches on that date, and run to 2100. ENSO hindcasts start on 1 May 1997--2021
with eight members and run for 12 months. The no-noise ablation (\SI{1}) is an earlier configuration without the
stochastic SST residual, with the 10-m wind closure acting on the full departure rather than on its low-pass part,
without the slow land-moisture injection and with its own flux adjustment. Two 30-year, two-member sensitivity controls
add the 80 leading modes of the SST-equation residual over 25--65$^\circ$ latitude, fitted as the tropical ones, at 0.5 and 0.7 of their fitted size (\ST{8}).

\subsection{Evaluation and data independence}
Climatologies are compared with ERA5 1979--2018 (the ocean's own climatology is that of 2000--2016), and the first two model years of each control are discarded. Errors are
area-weighted root-mean-square differences of annual or calendar-month means; Ni\~no-3.4 anomalies are monthly means over
5$^\circ$\,S--5$^\circ$\,N, 170--120$^\circ$\,W with the run's own seasonal cycle removed; El Ni\~no and La Ni\~na events
are five-month running means beyond $\pm0.5$\,K for at least five months. Blocking follows the two-dimensional index of
Davini et al.\cite{davini2012blocking}; tropical cyclones are 12-hourly MSLP minima over the sea south of 35$^\circ$\,N,
linked within 4.5$^\circ$ in 12\,h into tracks of at least three points. Standard errors of trends are adjusted for lag-1 autocorrelation. Hindcasts are scored raw, without bias correction, against ERA5, persistence and climatology; anomalies of the model and of ERA5 alike are taken from the 2000--2016 SST climatology, and lead $n$ is the $n$-th monthly mean, so lead 1 is May. The network was trained on ERA5 1979--2018 and the precipitation network on 1979--2016; the mixed-layer, sea-ice and noise coefficients, the 10-m wind relation, the diurnal diagnostic and the flux adjustment were fitted on 2000--2016 and the slow stress on 1979--2016; the wind bursts are resampled from 1979--2022; the energy balance
was constrained on 1979--2010 (temperature) and 2006--2020 (heat uptake) and the carbon cycle on 1959--2022. Free runs read
no ERA5 state after their initial condition, but four parts of the evaluation are not independent of ERA5 by design. The
mean climate is held near the ERA5 climatology by the 120-day relaxation and the flux adjustment, so its errors measure
how far the coupled model departs from that target. The network learned the weather of 1979--2018, so the daily
statistics of the controls are in-sample with respect to its training. The vertical structure of the forced response is
set by the forced-response layer and carried by the gauge transform. And the final configuration (the stochastic SST
residual, the low-pass wind observer and the strength of the slow stress) was chosen among candidates on their free-run
ENSO and on the May hindcasts, including those started in 2017--2021, which are outside every regression of the surface components but not outside this choice; the network's training also covers 2017--2018. The ocean-only cases of \ED{3} are driven by ERA5 fluxes that have felt the observed SST. River discharge (GRDC), the 1979--2022 warming pattern and the population and dam data were not used in any fit or choice; ORAS5 thermocline depths enter only the hindcast starts.

\section{Data availability}
ERA5 and ORAS5 are available from the Copernicus Climate Data Store, GRDC discharge from the Global Runoff Data Centre,
FLUXNET2015 from fluxnet.org, GLEAM from gleam.eu, the Mauna Loa \co\ record from NOAA, the ISIMIP3a dam and water-use data from the ISIMIP repository, and the CMIP7 forcing,
emissions and population data from input4MIPs.

\section{Code availability}
The \model\ code, the fitted coefficients, the run records and the scripts that produce the figures will be released on
publication.

\section{Competing interests}
The authors declare no competing interests.

\section{Additional information}
Supplementary Information is available for this paper. Correspondence and requests for materials should be addressed to
M.C.

\clearpage
\section*{Extended Data}
\setcounter{figure}{0}
\renewcommand{\figurename}{Extended Data Fig.}

\begin{figure}[!htbp]
\centering
\includegraphics[width=\textwidth]{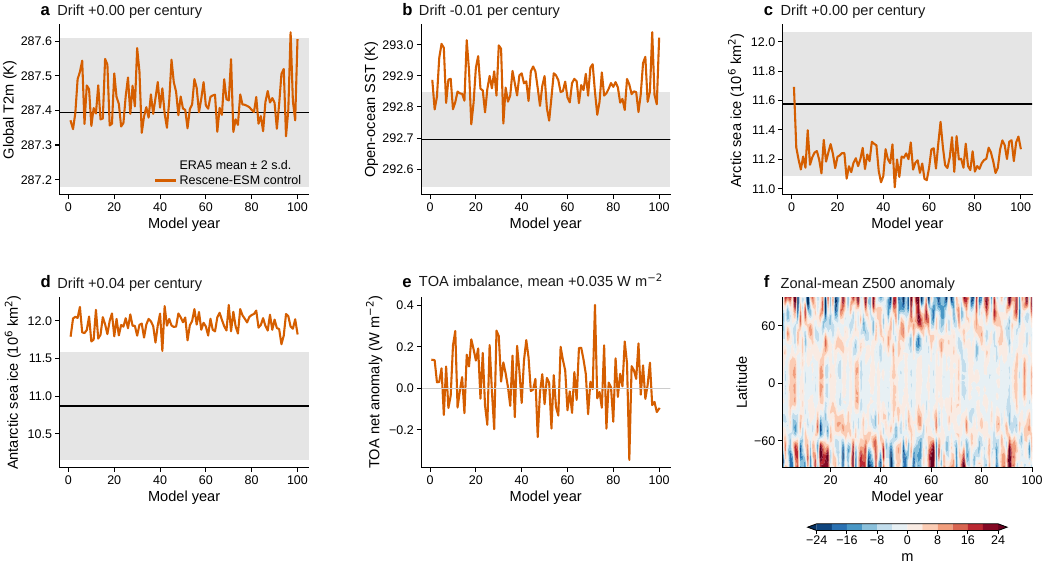}
\caption{\textbf{Stability of the 100-year control.} Annual means of \textbf{a}, global T2m; \textbf{b}, open-ocean SST;
\textbf{c}, Arctic and \textbf{d}, Antarctic sea-ice extent, with the ERA5 mean and $\pm2$ standard deviations (gray);
titles give the linear drift per century. \textbf{e}, Global top-of-atmosphere imbalance. \textbf{f}, Zonal-mean Z500
anomaly.}
\end{figure}

\begin{figure}[!htbp]
\centering
\includegraphics[width=\textwidth]{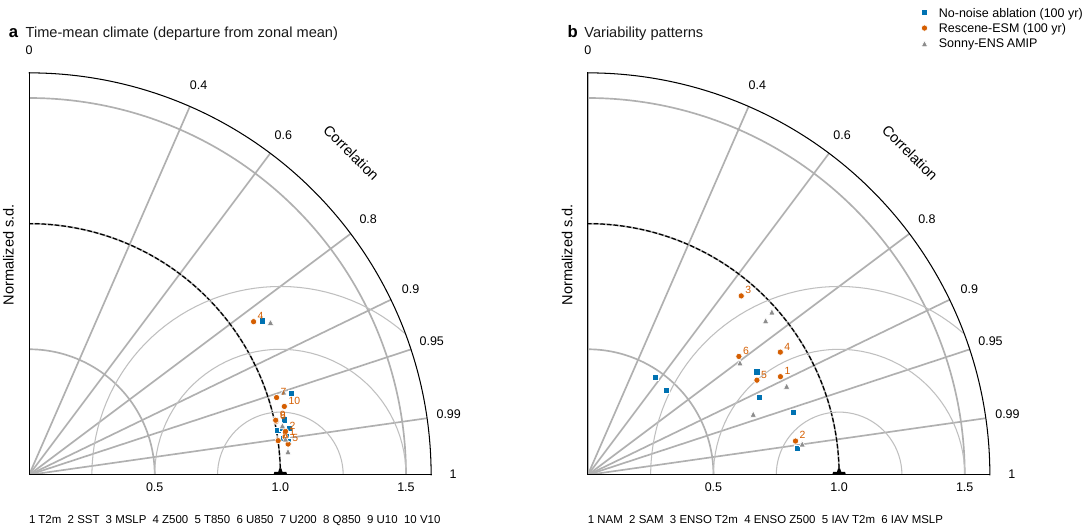}
\caption{\textbf{Taylor diagrams}\cite{taylor2001diagram}. \textbf{a}, Time-mean climate as departures from the zonal
mean. \textbf{b}, Variability patterns: the leading empirical orthogonal functions of extratropical MSLP (NAM, SAM), the
DJF regressions of T2m and Z500 on Ni\~no-3.4 (ENSO) and the interannual standard deviation of T2m and MSLP (IAV). Points:
\model\ and the no-noise ablation (100-year controls) and the same atmosphere driven by observed SST (AMIP, 1979--2022); ERA5 is the reference at correlation 1 and normalized standard deviation 1.}
\end{figure}

\begin{figure}[!htbp]
\centering
\includegraphics[width=\textwidth]{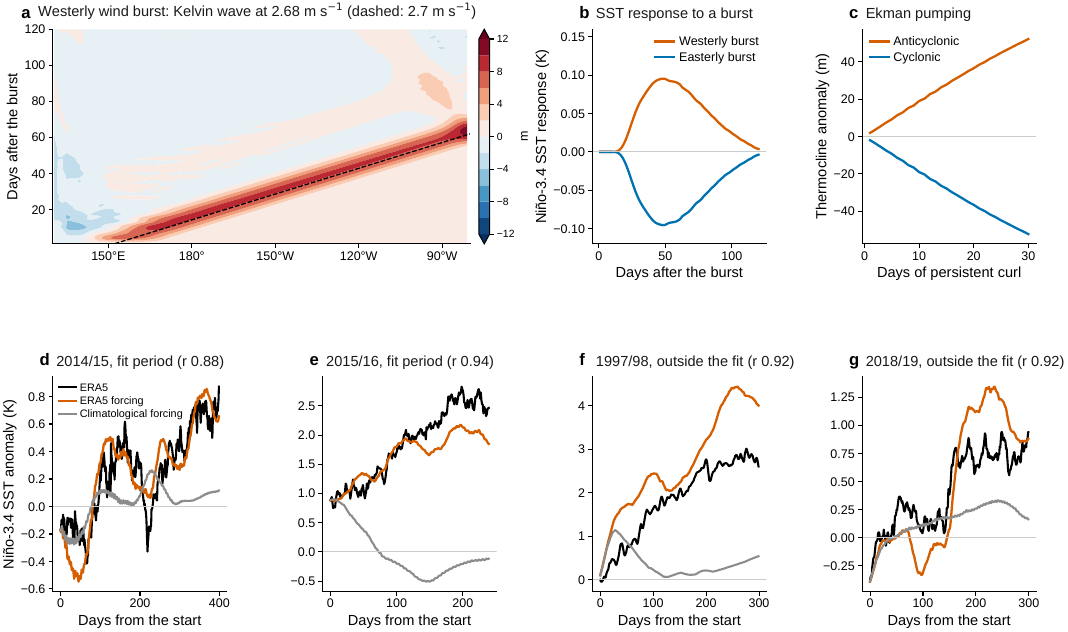}
\caption{\textbf{The ocean alone under prescribed winds}, the experiments of KIST-Ocean\cite{kim2026kistocean}.
\textbf{a}, Equatorial (2$^\circ$\,S--2$^\circ$\,N) thermocline anomaly after a 5-day westerly wind burst over
130--170$^\circ$\,E, 10$^\circ$\,S--10$^\circ$\,N; dashed, the Kelvin-wave speed of 2.7\,m\,s$^{-1}$. \textbf{b}, Ni\~no-3.4
SST response to the westerly and an equal easterly burst. \textbf{c}, Thermocline at the center of a persistent
cyclonic or anticyclonic wind-stress curl at 15.5$^\circ$\,N, 135.5$^\circ$\,E. \textbf{d--g}, Ni\~no-3.4 SST of the
ocean driven by ERA5 wind stress and surface fluxes and by climatological forcing, from the ERA5 state at the start, for
the 2014/15 and 2015/16 El Ni\~no (inside the 2000--2016 fit period) and the 1997/98 and 2018/19 El Ni\~no (outside it);
titles give the correlation with ERA5.}
\end{figure}

\begin{figure}[!htbp]
\centering
\includegraphics[width=\textwidth]{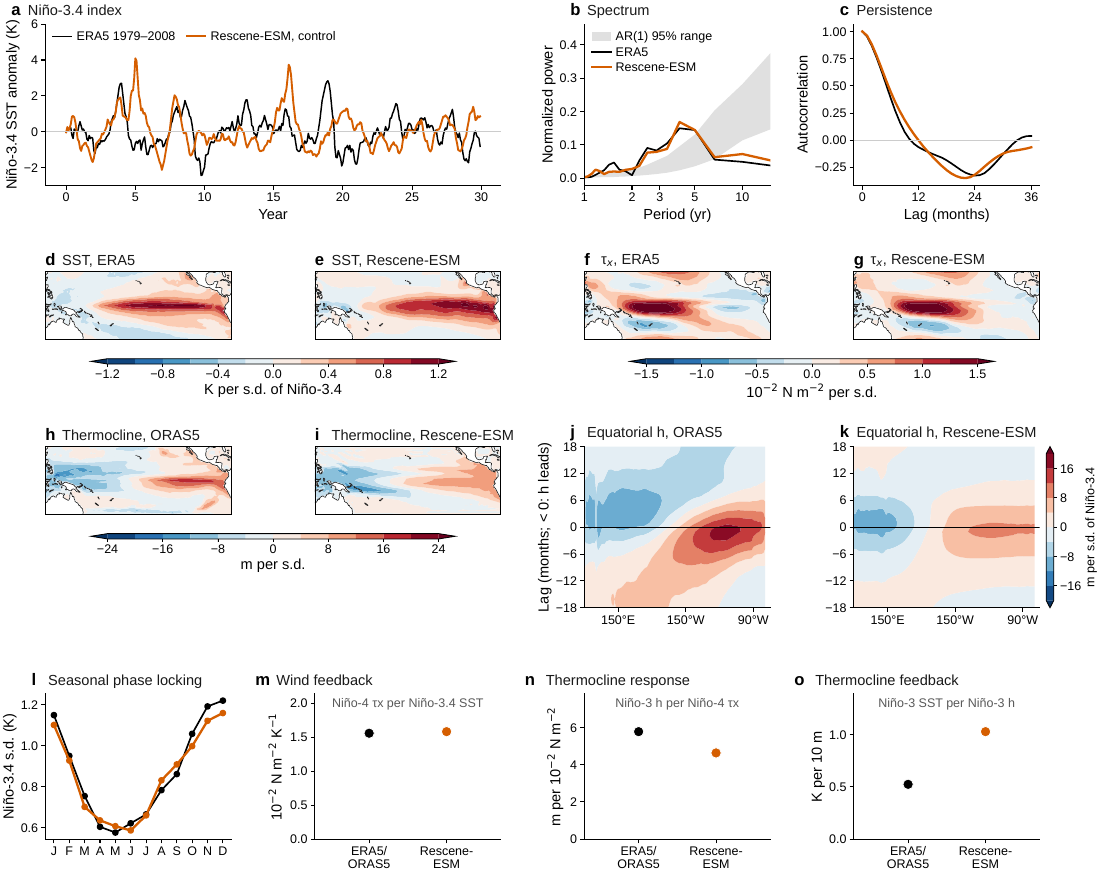}
\caption{\textbf{El Ni\~no in the coupled control.} \textbf{a}, Monthly Ni\~no-3.4 SST anomaly of the control (first 30
years shown) and of ERA5 1979--2008. \textbf{b}, Normalized spectra with the 95\% range of a fitted first-order
autoregressive process. \textbf{c}, Autocorrelation. \textbf{d--i}, Regression on Ni\~no-3.4 of SST (\textbf{d},
\textbf{e}), zonal wind stress (\textbf{f}, \textbf{g}) and thermocline depth (\textbf{h}, \textbf{i}; 20$^\circ$C
isotherm in ORAS5, the wave ocean's $h$ in the model) per standard deviation of the index. \textbf{j}, \textbf{k}, Lag
regression of equatorial thermocline depth on Ni\~no-3.4; negative lags: $h$ leads. \textbf{l}, Standard deviation of
Ni\~no-3.4 by calendar month. \textbf{m--o}, The links of the Bjerknes loop: Ni\~no-4 zonal stress per kelvin of
Ni\~no-3.4 SST, Ni\~no-3 thermocline depth per unit of Ni\~no-4 stress and Ni\~no-3 SST per 10\,m of Ni\~no-3
thermocline depth.}
\end{figure}

\begin{figure}[!htbp]
\centering
\includegraphics[width=\textwidth]{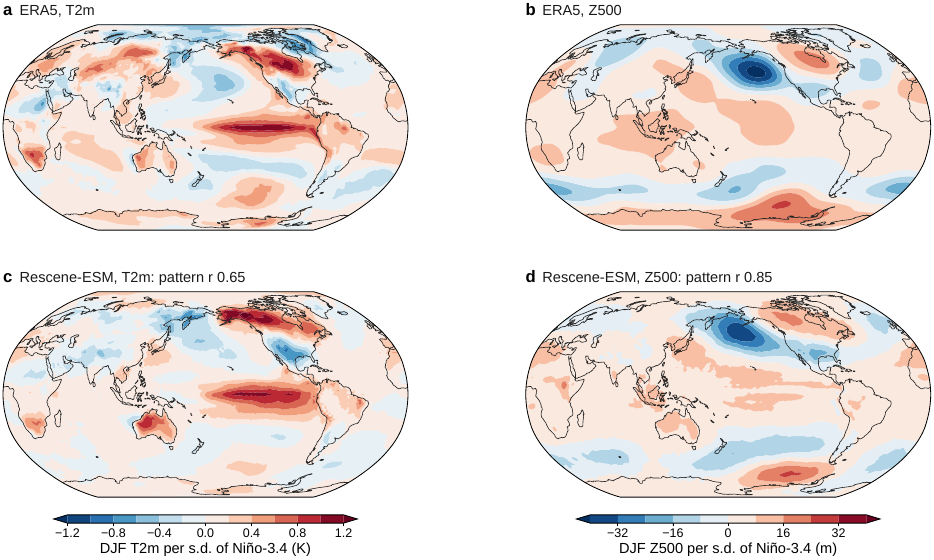}
\caption{\textbf{ENSO teleconnections.} December--February T2m (\textbf{a}, \textbf{c}) and Z500 (\textbf{b},
\textbf{d}) regressed on the Ni\~no-3.4 index in ERA5 and in the control. The model's patterns follow its own ENSO; no
statistical teleconnection is imposed.}
\end{figure}

\begin{figure}[!htbp]
\centering
\includegraphics[width=\textwidth]{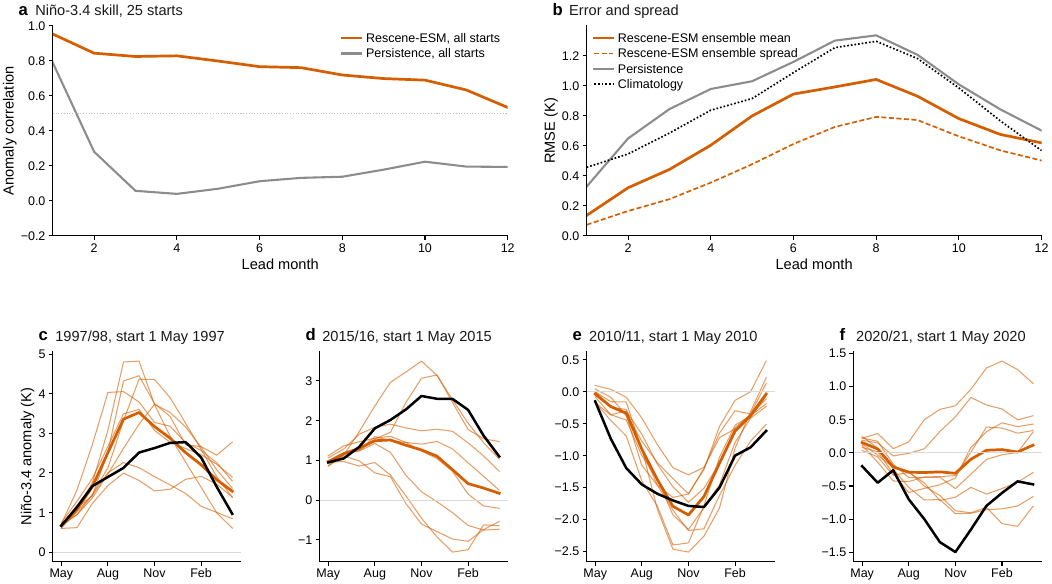}
\caption{\textbf{ENSO hindcasts.} Eight-member hindcasts started on 1 May of each year 1997--2021 (25 starts), scored raw
against ERA5. \textbf{a}, Anomaly correlation of the ensemble-mean Ni\~no-3.4 index and of persistence. \textbf{b}, RMSE
of the ensemble mean, of persistence and of climatology, and the ensemble spread. \textbf{c--f}, Members (thin) and
ensemble mean (thick) for the 1997/98 and 2015/16 El Ni\~no, the 2010/11 La Ni\~na and the 2020/21 La Ni\~na, with ERA5
(black). Starts in 2017--2021 are outside every regression of the surface components; the configuration was chosen partly on these hindcasts.}
\end{figure}

\begin{figure}[!htbp]
\centering
\includegraphics[width=\textwidth]{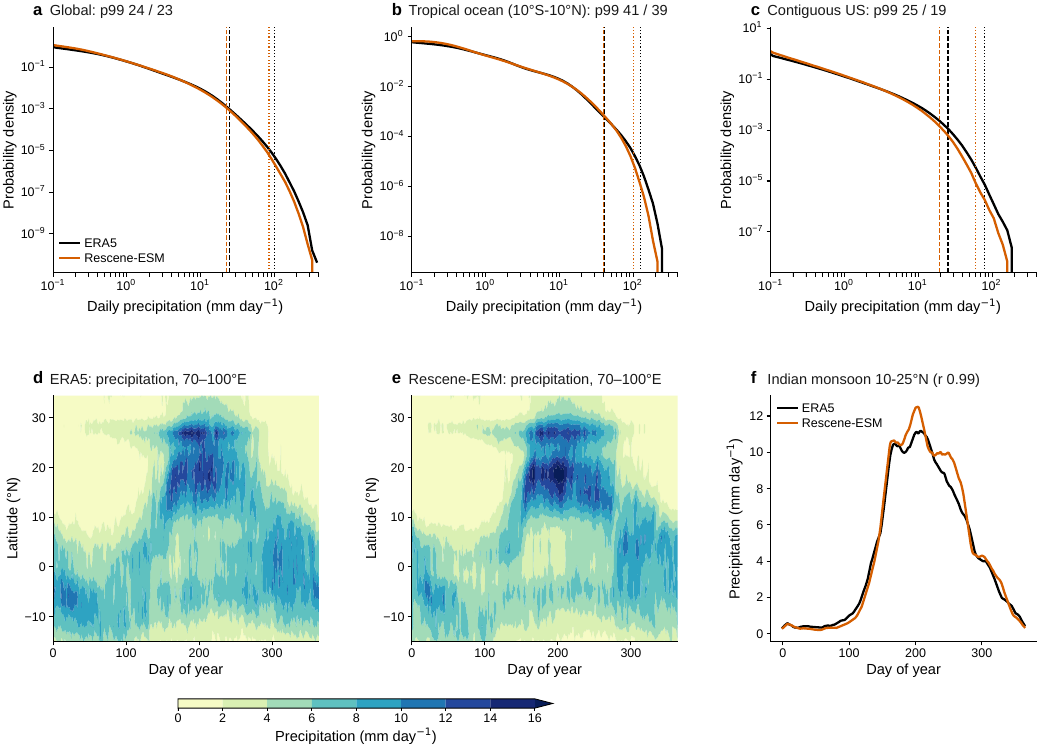}
\caption{\textbf{Daily precipitation and the Indian monsoon}, 40-year control against ERA5 1979--2018. \textbf{a--c},
Probability density of daily precipitation over the globe, the tropical ocean (10$^\circ$\,S--10$^\circ$\,N) and the
contiguous United States; dashed and dotted lines mark the 99th and 99.99th percentiles; titles give the 99th percentiles
of ERA5 and the model. \textbf{d}, \textbf{e}, Climatological daily precipitation over 70--100$^\circ$\,E by latitude and
day of year. \textbf{f}, Indian monsoon precipitation over 10--25$^\circ$\,N (15-day mean).}
\end{figure}

\begin{figure}[!htbp]
\centering
\includegraphics[width=\textwidth]{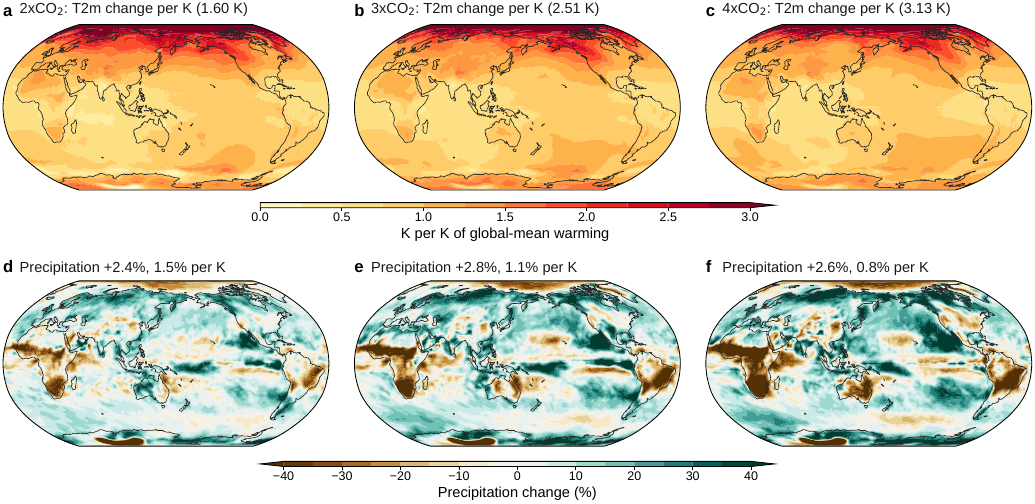}
\caption{\textbf{Patterns of the \co\ response.} Years 10--19 of abrupt 2$\times$, 3$\times$ and 4$\times$\co\ against
the fixed-\co\ control. \textbf{a--c}, T2m change per kelvin of global warming (titles give the global change).
\textbf{d--f}, Precipitation change (titles give the global change and the hydrological sensitivity).}
\end{figure}

\begin{figure}[!htbp]
\centering
\includegraphics[width=\textwidth]{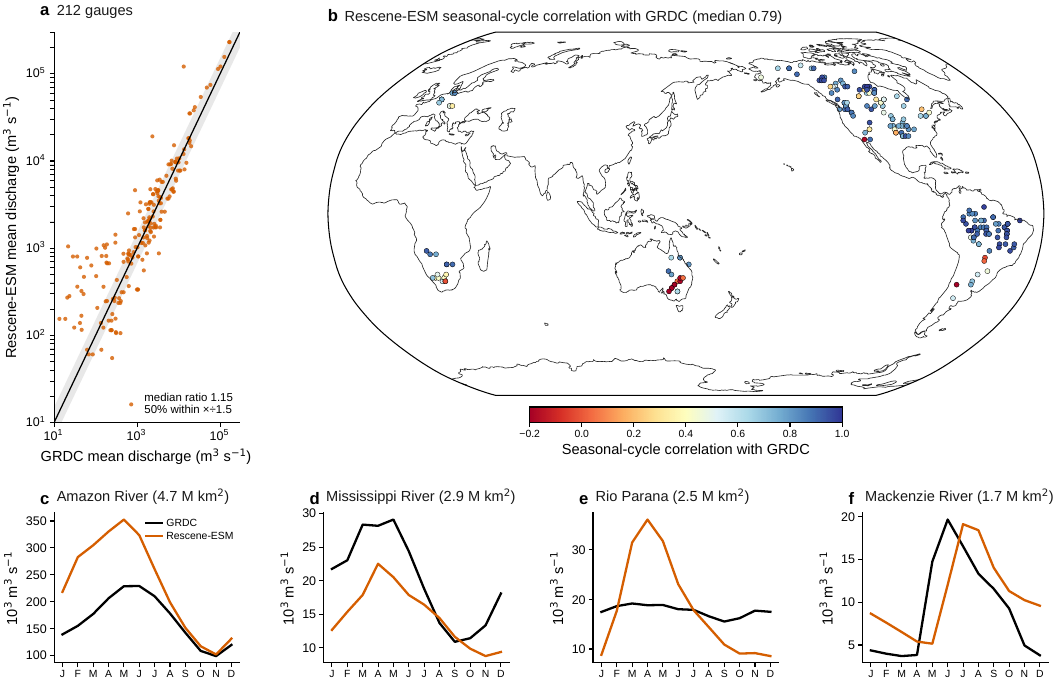}
\caption{\textbf{Rivers.} Discharge of the historical run at 212 GRDC gauges against observations (climatology from
1990). \textbf{a}, Mean discharge; shading, a factor of 1.5. \textbf{b}, Correlation of the seasonal cycle at each gauge.
\textbf{c--f}, Seasonal cycles of the four largest basins in the set at their main-stem gauges. No constant is fitted to
discharge.}
\end{figure}

\begin{figure}[!htbp]
\centering
\includegraphics[width=\textwidth]{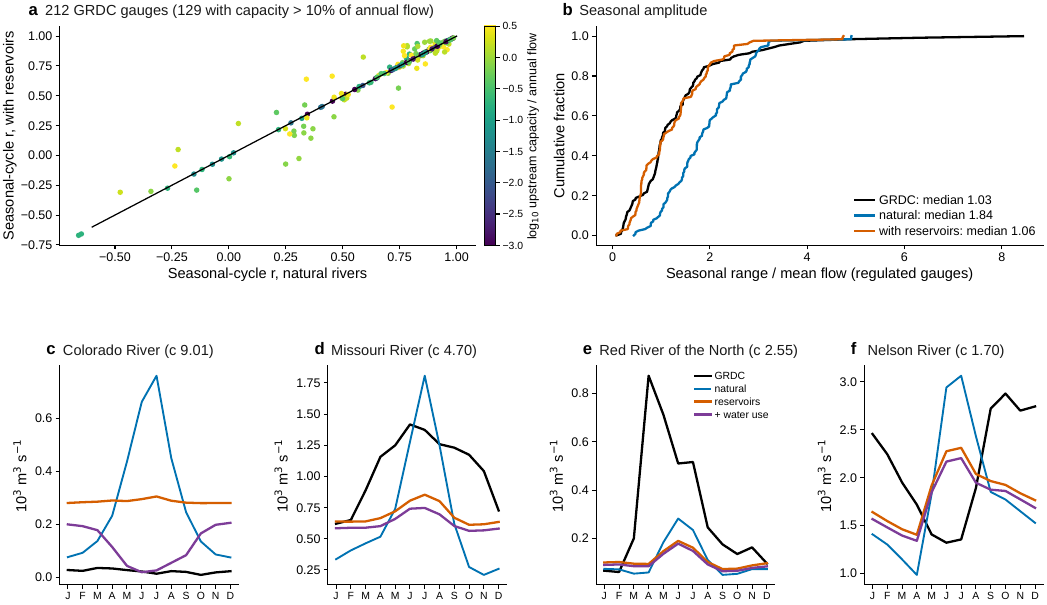}
\caption{\textbf{Reservoirs.} The historical runoff routed without and with the 7,302 ISIMIP3a dams (Hanasaki release
rule; no constant fitted to discharge), against GRDC climatologies from 1990. \textbf{a}, Seasonal-cycle correlation
without and with reservoirs, colored by the upstream capacity relative to the annual flow. \textbf{b}, Distribution of the
seasonal range relative to the mean flow at the 129 gauges with upstream capacity above 10\% of the annual flow.
\textbf{c--f}, Seasonal cycles of the most regulated large rivers; the purple lines add water consumption (sensitivity
run, Methods).}
\end{figure}

\begin{figure}[!htbp]
\centering
\includegraphics[width=\textwidth]{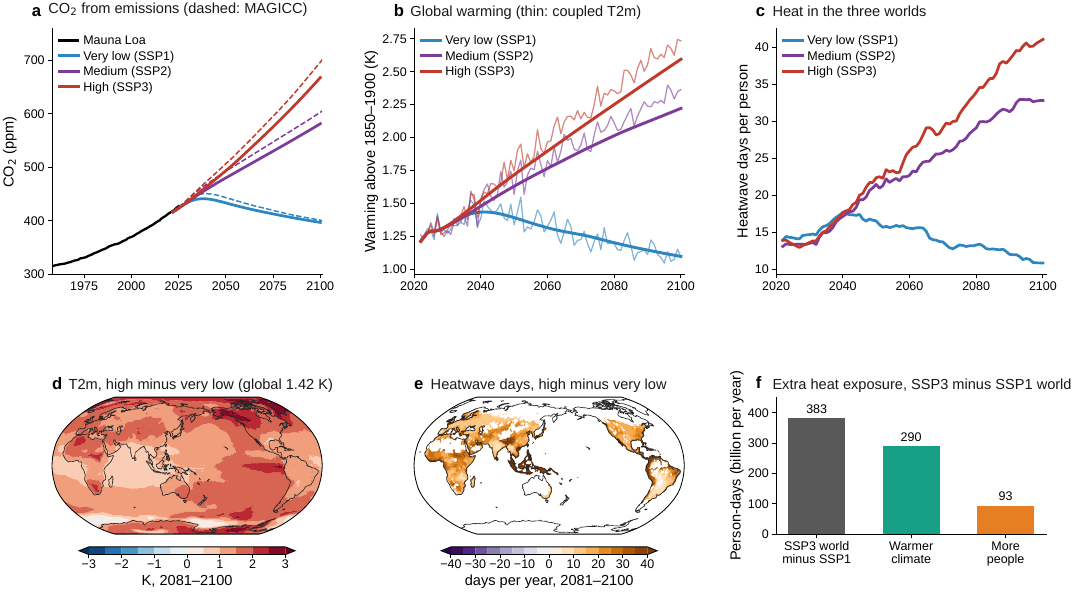}
\caption{\textbf{Emission scenarios.} CMIP7 very low, medium and high emissions from 2022, each with its own non-\co\ forcing and aerosol pattern and, for the exposure, its own population pathway (SSP1, SSP2, SSP3). \textbf{a},
Atmospheric \co\ from the carbon cycle, with Mauna Loa and each scenario's MAGICC concentration (dashed).
\textbf{b}, Global warming above 1850--1900 of the energy balance (thick) and the coupled model's global-mean T2m
(thin). \textbf{c}, Heatwave days per person (9-year running mean). \textbf{d}, \textbf{e}, T2m and heatwave days per
year in 2081--2100, high minus very low emissions. \textbf{f}, Heatwave person-days per year in 2081--2100, SSP3 world
minus SSP1 world, split into the part from the warmer climate and the part from the number of people (Methods).}
\end{figure}

\clearpage
\section*{Supplementary Information}
\setcounter{figure}{0}
\setcounter{table}{0}
\renewcommand{\figurename}{Supplementary Fig.}
\renewcommand{\tablename}{Supplementary Table}

\subsection*{Supplementary Tables}

\begin{table}[h]
\small
\caption{\textbf{From Rescene to \model.} Rescene is the predecessor that made a frozen weather model stable for a
century without interactive components. Metrics: 100-year controls against ERA5 1979--2018 (\model: model years 3--100).}
\begin{tabularx}{\textwidth}{>{\raggedright}p{3.6cm}>{\raggedright}X>{\raggedright\arraybackslash}X}
\toprule
 & Rescene & \model \\
\midrule
Atmosphere network & Sonny (frozen), deterministic 6-h step, 69 variables, with a 0.4-million-parameter wrapper &
Sonny-ENS-S (frozen), 20.8 million parameters, 12-h step with Monte Carlo dropout, 82 variables, SST as an input \\
Free-run stability & Slow clock: the large scales blended toward the fixed 1979--2018 calendar climatology every 6\,h
(weight 0.85 after 10 days) & Polar filter, stratospheric sponge and a 120-day relaxation toward the calendar climatology
plus the forced change \\
Day-to-day weather & Band-limited stochastic perturbation & The network's own dropout ensemble \\
Ocean and sea ice & Prescribed calendar climatology & Anomaly mixed layer, equatorial wave ocean, stochastic SST
residual, sea ice with thermodynamics and drift \\
Land and rivers & None & Soil-water bucket with groundwater and stomata, moisture flux into the boundary layer, rivers
and reservoirs \\
Forced change & None & Carbon cycle, two-layer energy balance, forced-response layer, gauge transform \\
People & None & Exposure recorder at 4,925 populated cells \\
\midrule
T2m / Z500 / MSLP RMSE & 1.69\,K / 33.3\,m / 0.91\,hPa & 0.64\,K / 21.1\,m / 4.16\,hPa \\
Global-mean annual T2m s.d. / ERA5 & 0.48 & 0.57 \\
Ni\~no-3.4 s.d. & None (climatological SST) & 0.91\,K (ERA5, 0.90\,K) \\
ENSO teleconnection pattern r, Z500 / T2m & None & 0.85 / 0.65 \\
1979--2022 warming trend / ERA5 & No forced change & 0.96 (first members of three runs) \\
Global T2m drift (K per century) & $-0.01$ & $+0.00$ \\
\bottomrule
\end{tabularx}
\end{table}

\begin{table}[h]
\small
\caption{\textbf{Experiments.} All runs use the same frozen network, coefficients and flux adjustment, except the
no-noise ablation, which has its own flux adjustment.}
\begin{tabularx}{\textwidth}{>{\raggedright}p{3.6cm}>{\raggedright}p{3.2cm}X}
\toprule
Experiment & Length, members & Forcing and use \tabularnewline
\midrule
Control & 100 yr, 2 & No forced change; mean climate, variability, ENSO \tabularnewline
1000-year control & 1000 yr, 1 & No forced change; annual-mean fields and monthly Ni\~no-3.4; long-term stability and
ENSO modulation (Supplementary Fig.~20) \tabularnewline
Daily control & 40 yr, 1 & As the control, with a daily record; precipitation, blocking, cyclones \tabularnewline
Coupling ladder & 4 $\times$ 30 yr, 2 & Atmosphere only; $+$ mixed layer and sea ice; $+$ wave ocean, slow stress and
bursts; $+$ stochastic SST residual (Fig.~2) \tabularnewline
Historical & 1979--2022, 3 runs $\times$ 2 & Observed \co\ and non-\co\ forcing \tabularnewline
Fixed-\co\ control & 20 yr, 2 & Reference of the \co\ experiments \tabularnewline
Abrupt 2$\times$, 3$\times$, 4$\times$\co & 20 yr each & Forced response \tabularnewline
1\%-per-year \co & 140 yr & Transient climate response \tabularnewline
SST perturbation & 5 yr & $+2$\,K open-ocean SST at the start \tabularnewline
Scenario & 2022--2100, 3 pathways & CMIP7 medium emissions scaled by SSP1/2/3 population \tabularnewline
Emission scenarios & 2022--2100, 2 & CMIP7 very low and high emissions, non-\co\ forcing and aerosol pattern;
SSP1 and SSP3 population (Extended Data Fig.~11) \tabularnewline
ENSO hindcasts & 25 starts $\times$ 8, 12 months & 1 May 1997--2021 \tabularnewline
Extratropical SST noise & 2 $\times$ 30 yr, 2 & The extratropical residual modes added at 0.5 and 0.7 of their fitted size
(Supplementary Table 8) \tabularnewline
Observed SST (AMIP) & 1979--2022, 8 & The same atmosphere driven by ERA5 SST and sea ice; reference in Fig.~2 and
Supplementary Tables 3 and 4 \tabularnewline
No-noise ablation & 100 yr, 2; 25 hindcasts & Without the stochastic SST residual, the low-pass wind observer and the slow
land-moisture injection; its own flux adjustment \tabularnewline
\bottomrule
\end{tabularx}
\end{table}

\begin{table}[h]
\small
\caption{\textbf{Coupling ladder.} 30-year, two-member controls with the couplings added cumulatively (model years
3--30; the full model: the same years of the 100-year control; Rescene: its 100-year run). Variability: standard
deviation of detrended annual means relative to ERA5 1979--2018 (median over ocean or land cells; global mean; sea-ice
concentration). ENSO teleconnection: pattern correlation with ERA5 of the DJF Z500 regression on each run's own
Ni\~no-3.4. AMIP: the same atmosphere driven by observed SST, 1979--2022. Rescene's value for single ocean cells reflects
its stochastic weather perturbation, which varies each cell while its global mean varies by half the observed amount.}
\setlength{\tabcolsep}{3pt}
\begin{tabularx}{\textwidth}{>{\raggedright\arraybackslash}Xccccccc}
\toprule
 & T2m / Z500 & \multicolumn{3}{c}{T2m variability} & Sea-ice & Ni\~no-3.4 & ENSO \\
 & RMSE & Ocean & Land & Global & variability & (K) & Z500 r \\
\midrule
Rescene & 1.69\,K / 33.3\,m & 1.19 & 0.63 & 0.48 & -- & -- & -- \\
Atmosphere only & 0.60\,K / 20.5\,m & 0.34 & 0.65 & 0.25 & 0.00 & 0.00 & -- \\
$+$ mixed layer and sea ice & 0.65\,K / 21.0\,m & 0.53 & 0.67 & 0.28 & 0.47 & 0.01 & -- \\
$+$ wave ocean, slow stress, bursts & 0.65\,K / 21.1\,m & 0.56 & 0.72 & 0.25 & 0.48 & 0.40 & 0.03 \\
$+$ SST noise & 0.65\,K / 21.0\,m & 0.68 & 0.73 & 0.45 & 0.50 & 0.85 & 0.65 \\
$+$ land moisture (\model) & 0.64\,K / 21.1\,m & 0.69 & 0.77 & 0.63 & 0.49 & 1.04 & 0.81 \\
AMIP (observed SST) & 0.57\,K / 20.7\,m & -- & -- & 0.72 & -- & -- & 0.81 \\
\bottomrule
\end{tabularx}
\end{table}

\begin{table}[h]
\small
\caption{\textbf{Coupled climate against ERA5}, 100-year control (model years 3--100). The same atmosphere driven by
observed SST over 1979--2022 (AMIP, eight members) and Rescene are shown for reference; AMIP teleconnections are by the measure of Fig.~2, on which \model\ has 0.84. River rows: historical runs.}
\begin{tabularx}{\textwidth}{Xcccc}
\toprule
Metric & ERA5 & \model & No-noise ablation & AMIP / Rescene \\
\midrule
T2m RMSE (K) & -- & 0.64 & 0.66 & 0.57 / 1.69 \\
Z500 RMSE (m) & -- & 21.1 & 21.1 & 20.7 / 33.3 \\
MSLP RMSE (hPa) & -- & 4.16 & 4.12 & 4.25 / 0.91 \\
U200 RMSE (m\,s$^{-1}$) & -- & 1.73 & 1.80 & -- \\
Land precipitation / evaporation RMSE (mm\,day$^{-1}$) & -- & 0.67 / 0.34 & 0.66 / 0.37 & -- \\
Global T2m drift (K per century) & 0 & $+0.00$ & $+0.02$ & -- \\
Arctic / Antarctic sea-ice area ($10^6$\,km$^2$) & 9.66 / 8.23 & 9.24 / 8.90 & 9.30 / 8.94 & -- \\
Ni\~no-3.4 s.d. (K) & 0.90 & 0.91 & 0.43 & -- \\
Ni\~no-3.4 spectral peak (yr) & 5.5 & 5 & -- & -- \\
Variance at 3--7 years (fraction) & 0.49 & 0.53 & 0.46 & -- \\
December / May s.d. ratio & 2.11 & 2.07 & 2.13 & -- \\
Ni\~no-3.4 skewness & $+0.51$ & $+1.12$ & $+0.22$ & -- \\
ENSO teleconnection pattern r, Z500 / T2m & 1 / 1 & 0.85 / 0.65 & 0.69 / 0.59 & 0.81 / 0.74 (AMIP) \\
NAM / SAM pattern r & 1 / 1 & 0.94 / 0.99 & -- & -- \\
River flow ratio, median / within $\times\div1.5$ & 1 / 100\% & 1.15 / 50\% & 1.17 / 48\% & -- \\
River seasonal-cycle r, median & 1 & 0.79 & 0.79 & -- \\
\bottomrule
\end{tabularx}
\end{table}

\begin{table}[h]
\small
\caption{\textbf{Ni\~no-3.4 hindcast skill} (ensemble mean of eight members, starts on 1 May). Values for 2017--2021 (five starts) are outside every regression of the surface components.}
\begin{tabularx}{\textwidth}{Xccccc}
\toprule
Lead (months) & 3 & 6 & 8 & 10 & 12 \\
\midrule
Anomaly correlation, 1997--2021 & 0.83 & 0.77 & 0.72 & 0.69 & 0.53 \\
RMSE (K), 1997--2021 & 0.44 & 0.94 & 1.04 & 0.78 & 0.62 \\
Ensemble spread (K) & 0.24 & 0.61 & 0.79 & 0.66 & 0.50 \\
Persistence RMSE (K) & 0.84 & 1.16 & 1.34 & 1.01 & 0.70 \\
Climatology RMSE (K) & 0.68 & 1.09 & 1.30 & 0.99 & 0.57 \\
Anomaly correlation, 2017--2021 & 0.48 & 0.72 & 0.79 & 0.84 & 0.94 \\
RMSE (K), 2017--2021 & 0.33 & 0.85 & 0.95 & 0.52 & 0.56 \\
No-noise ablation: RMSE / spread (K), 1997--2021 & 0.45 / 0.13 & 1.02 / 0.30 & 1.16 / 0.40 & 0.85 / 0.31 & 0.65 / 0.25 \\
\bottomrule
\end{tabularx}
\end{table}

\begin{table}[h]
\small
\caption{\textbf{Diagnostics in the format of AI coupled-model studies} (ACE2-SOM, SamudrACE, DLESyM, ACE2-NEMO).
Control runs; ERA5 1979--2018 (daily) or 1979--2022 (ENSO). ED, Extended Data Fig.; SI, Supplementary Fig.}
\begin{tabularx}{\textwidth}{Xccc}
\toprule
Diagnostic & ERA5 & \model & Figure \\
\midrule
El Ni\~no / La Ni\~na events per 40 years & 10.0 / 10.9 & 8.2 / 11.0 & SI 9 \\
El Ni\~no mean / maximum peak (K) & 1.76 / 2.84 & 1.99 / 4.37 & SI 9 \\
Pacific tripole index s.d. / 13-year low-pass s.d. (K) & 0.78 / 0.11 & 0.75 / 0.13 & SI 10 \\
Gulf Stream / Kuroshio Extension cell SST s.d. (K) & 0.89 / 0.80 & 0.35 / 0.37 & SI 10 \\
Sea-ice concentration variability, NH / SH & 0.125 / 0.137 & 0.055 / 0.065 & SI 11 \\
Seasonal-cycle RMSE by latitude, SST / T2m (K) & -- & 0.26 / 0.50 & SI 4 \\
DJF blocking pattern r / frequency ratio & 1 / 1 & 0.72 / 0.99 & SI 12 \\
Western North Pacific cyclones per year, $<$994 / 985 / 975\,hPa & 15.3 / 8.1 / 4.1 & 17.3 / 7.7 / 2.9 & SI 13 \\
Daily precipitation mean / 99th / 99.99th percentile (mm\,day$^{-1}$) & 2.40 / 24.4 / 101 & 2.28 / 22.6 / 85 & ED 7 \\
Indian monsoon annual-cycle r / June--September ratio / onset day & 1 / 1 / 149 & 0.99 / 1.09 / 150 & ED 7 \\
ENSO precipitation regression r / amplitude ratio (30$^\circ$\,S--30$^\circ$\,N) & 1 / 1 & 0.79 / 0.79 & SI 8 \\
\bottomrule
\end{tabularx}
\end{table}

\begin{table}[h]
\small
\caption{\textbf{Forced response and futures.} Abrupt experiments: years 10--19 against the fixed-\co\ control.
Futures: 2081--2100, CMIP7 medium emissions scaled by population, and each scenario's own emissions with its own
population.}
\begin{tabularx}{\textwidth}{Xccc}
\toprule
 & 2$\times$\co & 3$\times$\co & 4$\times$\co \\
\midrule
Global T2m change (K) & 1.60 & 2.51 & 3.13 \\
Global precipitation change (\%) & $+2.4$ & $+2.8$ & $+2.6$ \\
Hydrological sensitivity (\% K$^{-1}$) & 1.52 & 1.11 & 0.85 \\
Land / ocean warming ratio & 1.30 & 1.24 & 1.21 \\
Arctic (north of 60$^\circ$\,N) amplification & 2.40 & 2.14 & 2.11 \\
\midrule
 & SSP1 population & SSP2 population & SSP3 population \\
\midrule
Population in 2100 (billion) & 8.2 & 10.1 & 13.2 \\
\co\ in 2100 (ppm; MAGICC, 603) & 565 & 581 & 607 \\
Warming above 1850--1900 (K) & 2.08 & 2.12 & 2.18 \\
Heatwave exposure (billion person-days per year) & 265 & 323 & 410 \\
River-flood days per person (1995--2014, 1.78) & 2.78 & 2.28 & 2.39 \\
Drought months per person (1995--2014, 1.19) & 1.18 & 1.26 & 1.34 \\
\midrule
 & Very low, SSP1 & Medium, SSP2 & High, SSP3 \\
\midrule
\co\ in 2100 (ppm) & 397 & 581 & 668 \\
\co\ in 2100, MAGICC (ppm) & 401 & 603 & 697 \\
Warming above 1850--1900 (K) & 1.14 & 2.12 & 2.43 \\
Heatwave days per person & 12.0 & 31.7 & 38.3 \\
Heatwave exposure (billion person-days per year) & 104 & 323 & 487 \\
\bottomrule
\end{tabularx}
\end{table}

\begin{table}[h]
\small
\caption{\textbf{Extratropical SST noise (sensitivity).} 30-year controls, first member, model years 3--30. The 80
leading modes of the SST-equation residual over 25--65$^\circ$ latitude, fitted as the 20 tropical ones, are added to
them at 0.5 and 0.7 of their fitted size. Boundary currents: s.d. of monthly SST anomalies averaged over the cells of
the Gulf Stream (35--45$^\circ$\,N, 75--45$^\circ$\,W) and Kuroshio (30--40$^\circ$\,N, 140--170$^\circ$\,E) Extension
boxes, and of the box means. Ratios: median over cells of the model's s.d. over ERA5's (1979--2022); mean errors against
ERA5 1979--2018.}
\begin{tabularx}{\textwidth}{Xcccc}
\toprule
 & ERA5 & \model & Noise $\times0.5$ & Noise $\times0.7$ \\
\midrule
Gulf Stream / Kuroshio SST s.d., cells (K) & 0.89 / 0.80 & 0.32 / 0.35 & 0.45 / 0.52 & 0.56 / 0.57 \\
Gulf Stream / Kuroshio SST s.d., box means (K) & 0.50 / 0.52 & 0.23 / 0.23 & 0.29 / 0.37 & 0.37 / 0.37 \\
Extratropical (30--60$^\circ$) monthly SST s.d. ratio & 1 & 0.52 & 0.65 & 0.72 \\
Interannual T2m s.d. ratio, ocean / land & 1 / 1 & 0.68 / 0.76 & 0.72 / 0.78 & 0.70 / 0.75 \\
Ni\~no-3.4 s.d. (K) & 0.90 & 1.04 & 0.99 & 0.86 \\
T2m / SST RMSE (K) & -- & 0.64 / 0.37 & 0.63 / 0.37 & 0.64 / 0.37 \\
Sea-ice concentration s.d. & 0.131 & 0.060 & 0.060 & 0.059 \\
Global T2m trend (K per century) & -- & $+0.00$ & $+0.12$ & $+0.20$ \\
\bottomrule
\end{tabularx}
\end{table}

\begin{table}[h]
\small
\caption{\textbf{Heat exposure on the metrics of multi-model studies.} Warm-spell index as in Dash et al. (npj Nat.
Hazards, 2025; factor of 2--3 over 2021--2050 in 19 CMIP6 models), over the populated cells; in 1991--2020 the model has
4.27 and ERA5 5.00 days per year. Exposure is the product of the period-mean heatwave days and population, in billion person-days per year (hence 486 against the 487 of Supplementary Table 7). Each rise is given with its split into climate /
population / their product, in percent. India, China, Pakistan and Bangladesh hold 73\% of the people exposed at 3\,K of
warming in Kong et al. (Lancet Planet. Health, 2026); on the 1.5$^\circ$ grid, ERA5 exceeds the wet-bulb limit of 30.6$^\circ$C quoted there on at most 0.3 days per year in any populated cell.}
\begin{tabularx}{\textwidth}{Xccc}
\toprule
 & Very low, SSP1 & Medium, SSP2 & High, SSP3 \\
\midrule
\multicolumn{4}{l}{\textit{Warm-spell index}} \\
Days per year, 2023--2050 & 9.4 & 9.9 & 10.2 \\
Ratio to 1991--2020, area-weighted & 2.20 & 2.33 & 2.38 \\
Ratio to 1991--2020, population-weighted & 2.02 & 2.01 & 2.10 \\
\midrule
\multicolumn{4}{l}{\textit{Heatwave exposure, 2023--2050, against 57.7 in 1991--2020}} \\
Exposure & 139 & 144 & 150 \\
World & 82 (64 / 19 / 17) & 86 (63 / 19 / 18) & 92 (62 / 19 / 19) \\
Asia & 51 (68 / 17 / 15) & 51 (67 / 17 / 16) & 55 (66 / 17 / 17) \\
Africa & 14 (34 / 35 / 31) & 16 (32 / 35 / 32) & 17 (31 / 36 / 33) \\
Europe & 6.0 (97 / 1 / 2) & 5.8 (98 / 0 / 1) & 6.1 (98 / 0 / 2) \\
North America & 6.8 (67 / 17 / 16) & 7.9 (67 / 15 / 18) & 8.7 (69 / 14 / 17) \\
South America & 3.1 (65 / 20 / 15) & 4.7 (68 / 14 / 18) & 4.4 (65 / 16 / 18) \\
Oceania & 0.5 (54 / 21 / 25) & 0.7 (55 / 18 / 27) & 0.7 (54 / 17 / 29) \\
\midrule
\multicolumn{4}{l}{\textit{Heatwave exposure, 2081--2100, against 39.3 in 1986--2005}} \\
Exposure & 104 & 323 & 486 \\
World & 64 (66 / 18 / 16) & 284 (55 / 7 / 38) & 447 (44 / 7 / 49) \\
Asia & 40 (81 / 8 / 11) & 150 (69 / 4 / 26) & 222 (56 / 6 / 38) \\
Africa & 14 (14 / 53 / 33) & 82 (19 / 12 / 69) & 142 (14 / 10 / 75) \\
Europe & 3.8 (115 / $-$12 / $-$3) & 13 (103 / $-$2 / $-$1) & 19 (92 / 0 / 8) \\
North America & 5.7 (55 / 23 / 22) & 23 (61 / 7 / 32) & 35 (53 / 6 / 41) \\
South America & 1.2 (60 / 27 / 13) & 14 (69 / 4 / 27) & 24 (58 / 4 / 38) \\
Oceania & 0.5 (36 / 26 / 38) & 2.9 (40 / 6 / 54) & 4.7 (34 / 4 / 62) \\
\midrule
\multicolumn{4}{l}{\textit{India, China, Pakistan and Bangladesh, 2081--2100}} \\
Share of people (\%) & 31 & 29 & 29 \\
Share of heatwave person-days (\%) & 35 & 25 & 23 \\
Heatwave days per person, there / elsewhere & 13.8 / 11.2 & 27.7 / 33.4 & 30.6 / 41.4 \\
\bottomrule
\end{tabularx}
\end{table}

\clearpage
\subsection*{Supplementary Figures}

\begin{figure}[!htbp]
\centering
\includegraphics[width=\textwidth]{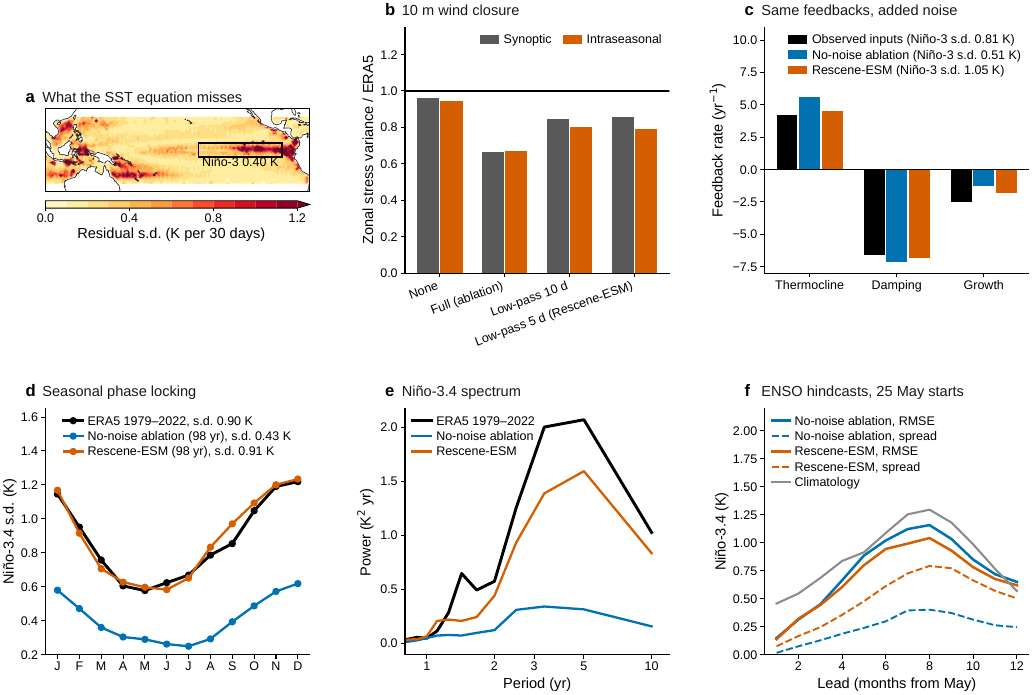}
\caption{\textbf{What sets the coupled ENSO.} \textbf{a}, Standard deviation of the 30-day residual of the mixed-layer SST equation on ERA5 2000--2016, the stochastic tendency of the model. \textbf{b}, Variance of the equatorial Pacific zonal
stress at synoptic and intraseasonal scales relative to ERA5, in 120-day free runs of the network without a 10-m wind
closure, with the closure on the full departure (ablation) and with it acting on the departure's running mean over
10 days and over 5 days (\model). \textbf{c}, Ni\~no-3 feedback rates of the SST equation (thermocline, damping and net growth) and the
Ni\~no-3 standard deviation with observed inputs, in the no-noise ablation and in \model. \textbf{d}, Seasonal phase
locking, \textbf{e}, spectrum and \textbf{f}, hindcast error and spread of the ablation and of \model.}
\end{figure}

\begin{figure}[!htbp]
\centering
\includegraphics[width=\textwidth]{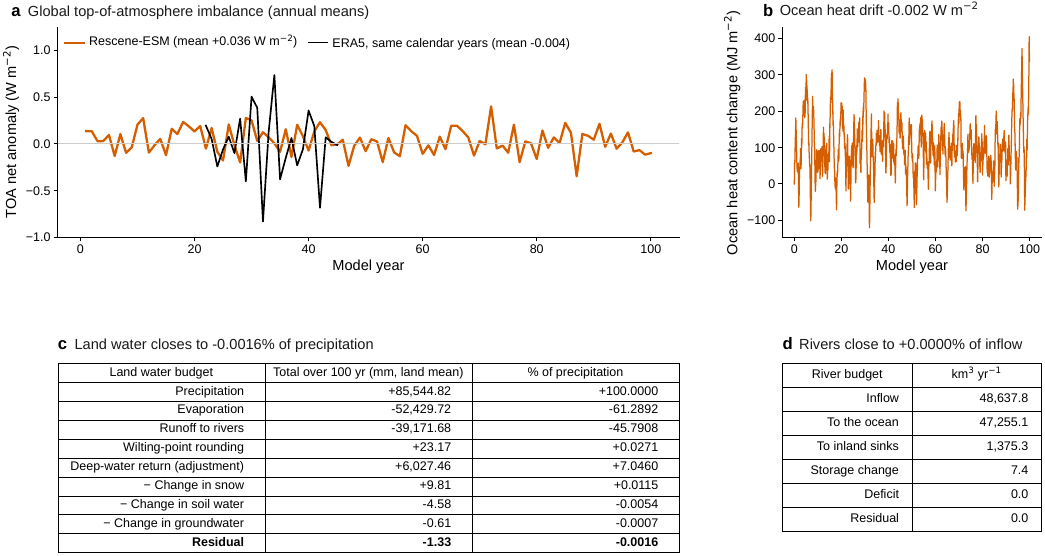}
\caption{\textbf{Energy and water budgets of the 100-year control.} \textbf{a}, Global top-of-atmosphere imbalance
(annual means) and ERA5 for the same calendar years. \textbf{b}, Ocean heat content change. \textbf{c}, Land water budget
over 100 years; every term is a ledger of the run. \textbf{d}, River budget.}
\end{figure}

\begin{figure}[!htbp]
\centering
\includegraphics[width=\textwidth]{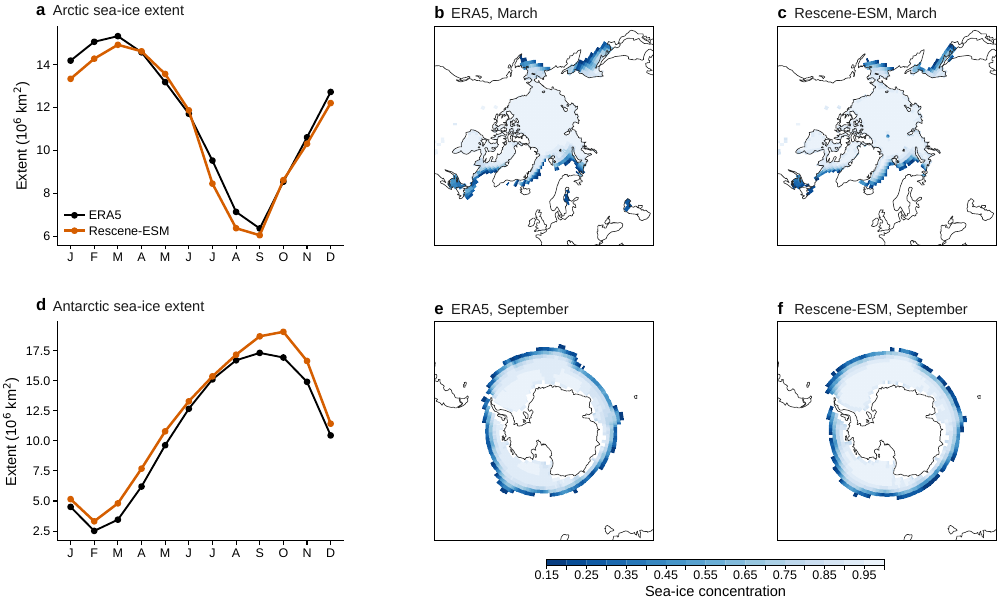}
\caption{\textbf{Sea ice.} \textbf{a}, \textbf{d}, Seasonal cycle of Arctic and Antarctic sea-ice extent (concentration
above 0.15). \textbf{b}, \textbf{c}, \textbf{e}, \textbf{f}, Sea-ice concentration at the seasonal maxima, March in the
Arctic and September in the Antarctic.}
\end{figure}

\begin{figure}[!htbp]
\centering
\includegraphics[width=\textwidth]{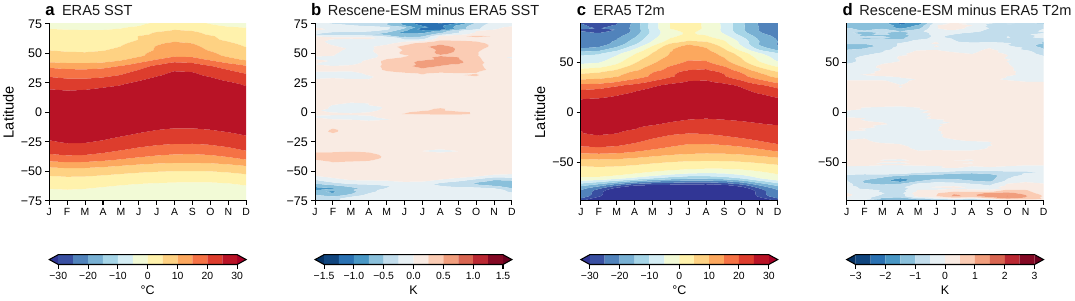}
\caption{\textbf{Seasonal cycle by latitude.} Zonal-mean calendar-month SST (\textbf{a}) and T2m (\textbf{c}) of ERA5 and
the departure of the control (\textbf{b}, \textbf{d}).}
\end{figure}

\begin{figure}[!htbp]
\centering
\includegraphics[width=\textwidth]{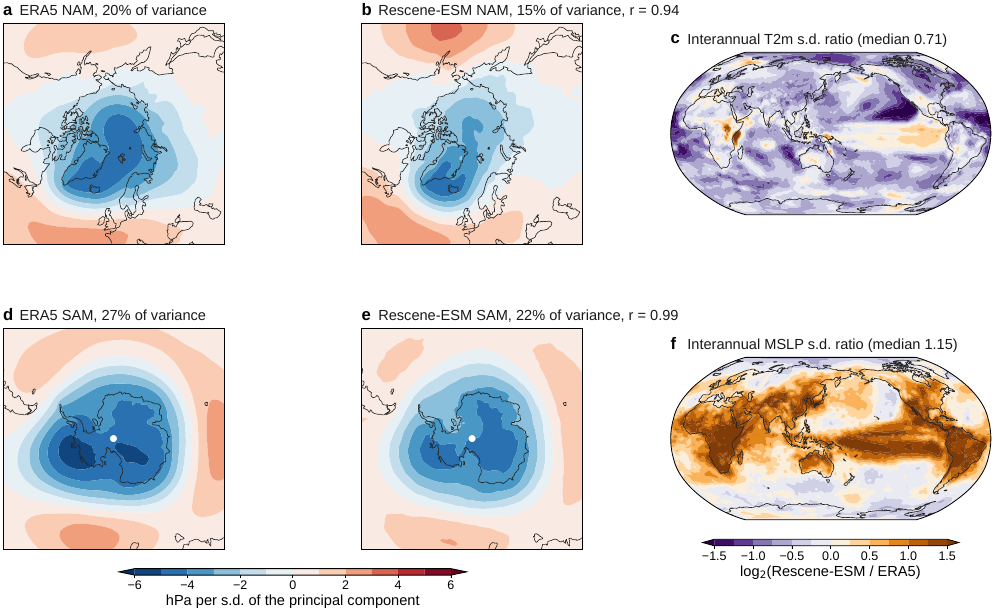}
\caption{\textbf{Annular modes and interannual variability.} \textbf{a}, \textbf{b}, \textbf{d}, \textbf{e}, Leading
empirical orthogonal function of monthly MSLP anomalies over the extratropical Northern (NAM) and Southern (SAM)
Hemisphere, regressed on the principal component. \textbf{c}, \textbf{f}, Ratio of the interannual standard deviation of
T2m and MSLP to ERA5, mapped as $\log_2$; titles give the median ratio.}
\end{figure}

\begin{figure}[!htbp]
\centering
\includegraphics[width=\textwidth]{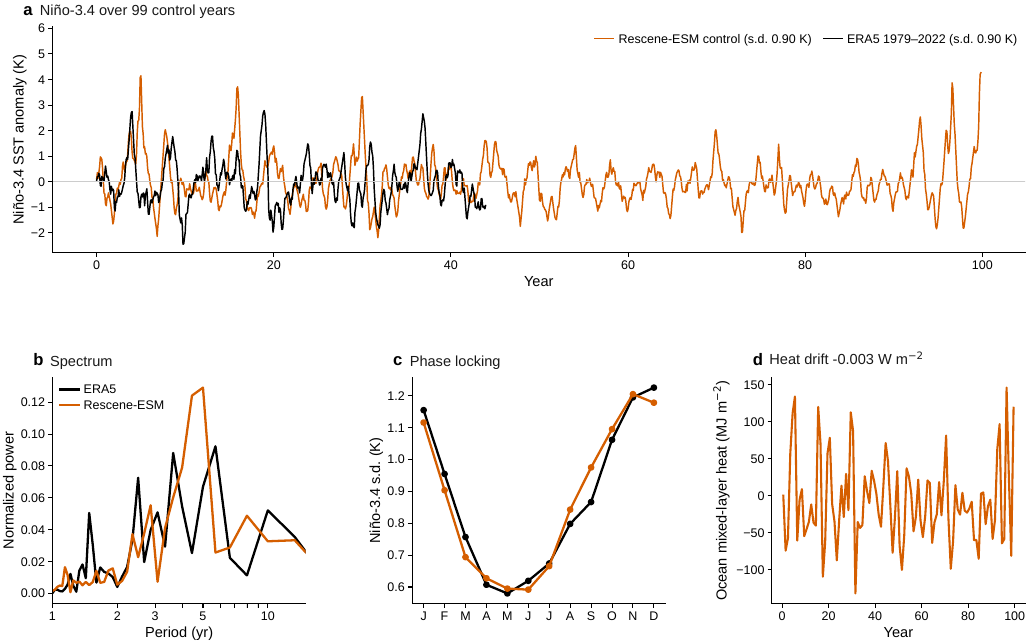}
\caption{\textbf{ENSO and heat over the full control.} \textbf{a}, Monthly Ni\~no-3.4 anomaly over the 100 years, with
ERA5 1979--2022. \textbf{b}, Spectra (40-year Hann windows, half overlapping). \textbf{c}, Standard deviation by calendar
month. \textbf{d}, Annual heat content of the ocean mixed layer; the title gives its linear drift.}
\end{figure}

\begin{figure}[!htbp]
\centering
\includegraphics[width=\textwidth]{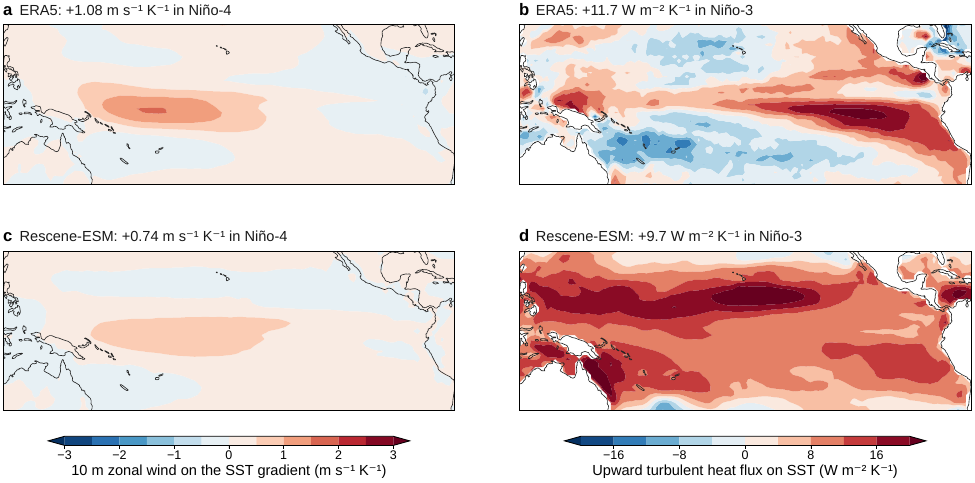}
\caption{\textbf{ENSO feedbacks.} \textbf{a}, \textbf{c}, Regression of the 10-m zonal wind on the zonal SST gradient
(the Bjerknes wind feedback; titles give the Ni\~no-4 mean). \textbf{b}, \textbf{d}, Regression of the upward turbulent
heat flux on SST (the heat-flux damping; titles give the Ni\~no-3 mean).}
\end{figure}

\begin{figure}[!htbp]
\centering
\includegraphics[width=\textwidth]{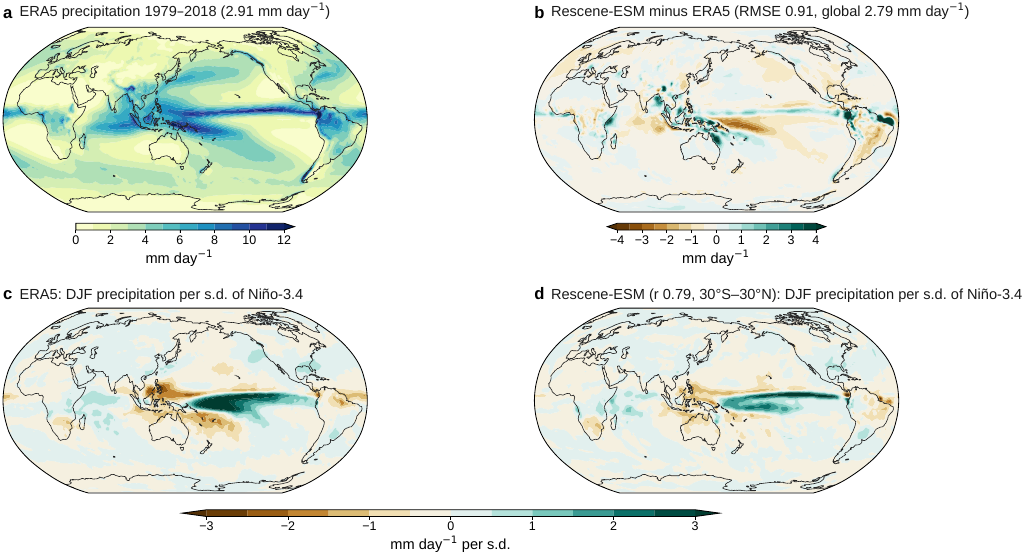}
\caption{\textbf{Mean precipitation and its ENSO response.} \textbf{a}, ERA5 annual-mean precipitation, 1979--2018.
\textbf{b}, Departure of the 40-year control. \textbf{c}, \textbf{d}, December--February precipitation regressed on
Ni\~no-3.4 in ERA5 and in the control; the title gives the pattern correlation over 30$^\circ$\,S--30$^\circ$\,N.}
\end{figure}

\begin{figure}[!htbp]
\centering
\includegraphics[width=\textwidth]{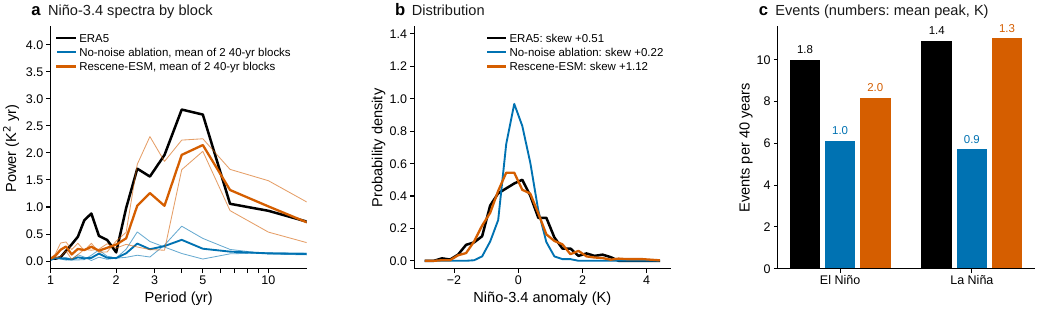}
\caption{\textbf{ENSO events.} \textbf{a}, Ni\~no-3.4 spectra of 40-year blocks (thin) and their mean (thick).
\textbf{b}, Distribution of monthly Ni\~no-3.4 anomalies. \textbf{c}, El Ni\~no and La Ni\~na events per 40 years
(five-month running mean beyond $\pm0.5$\,K for at least five months); numbers give the mean peak.}
\end{figure}

\begin{figure}[!htbp]
\centering
\includegraphics[width=\textwidth]{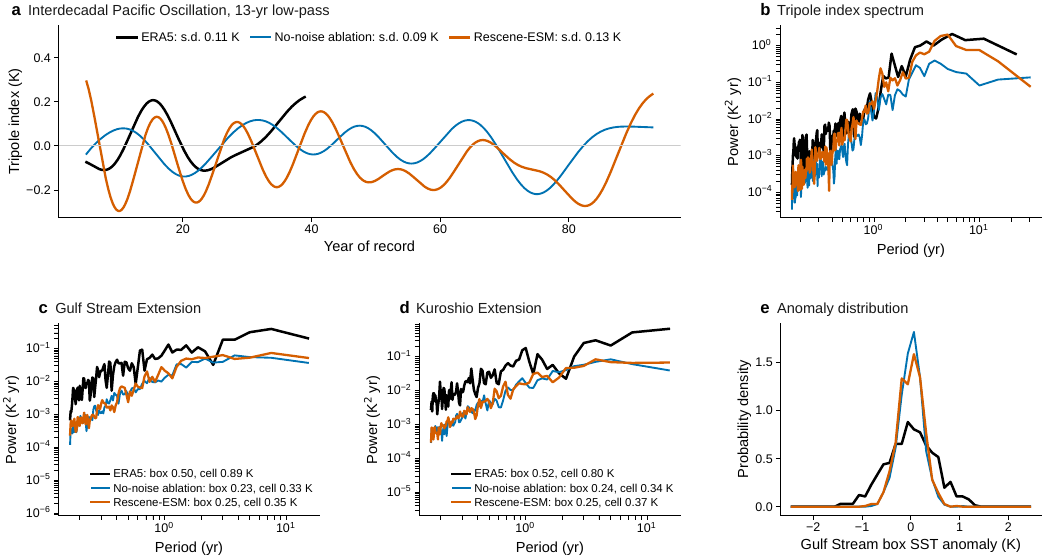}
\caption{\textbf{Decadal variability and the western boundary currents.} \textbf{a}, 13-year low-pass Pacific tripole
index. \textbf{b}, Its spectrum. \textbf{c}, \textbf{d}, Spectra of SST anomalies averaged over the Gulf Stream and
Kuroshio Extensions; legends give the standard deviation of the box mean and of individual cells. \textbf{e},
Distribution of the Gulf Stream box anomaly.}
\end{figure}

\begin{figure}[!htbp]
\centering
\includegraphics[width=\textwidth]{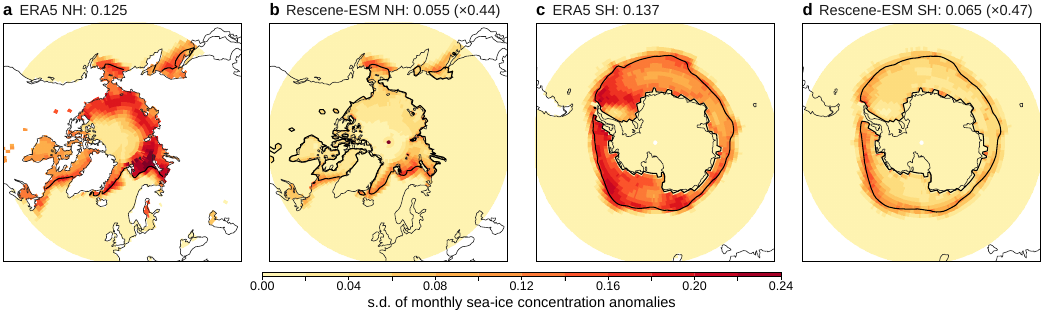}
\caption{\textbf{Sea-ice variability.} Standard deviation of monthly sea-ice concentration anomalies in ERA5 and in the
control, Northern (\textbf{a}, \textbf{b}) and Southern (\textbf{c}, \textbf{d}) Hemispheres; titles give the area mean
and its ratio to ERA5. Contours: climatological ice edge.}
\end{figure}

\begin{figure}[!htbp]
\centering
\includegraphics[width=\textwidth]{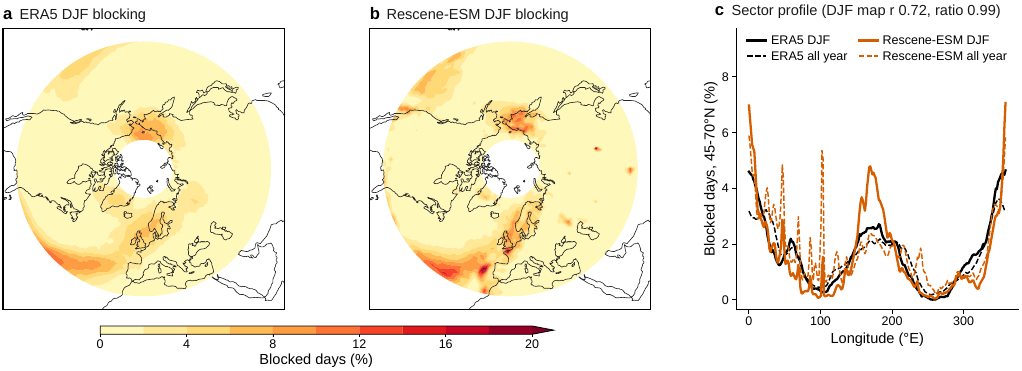}
\caption{\textbf{Northern Hemisphere blocking}, two-dimensional index on daily Z500, 40-year control against ERA5
1979--2018. \textbf{a}, \textbf{b}, Frequency of blocked days in December--February. \textbf{c}, Frequency over
45--70$^\circ$\,N by longitude, winter and the whole year.}
\end{figure}

\begin{figure}[!htbp]
\centering
\includegraphics[width=\textwidth]{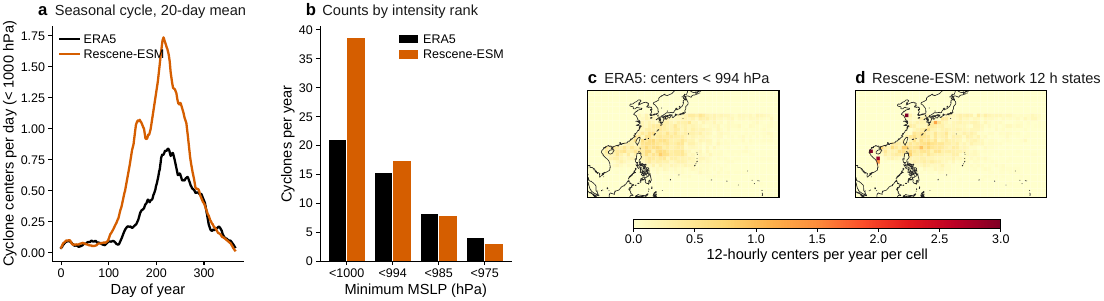}
\caption{\textbf{Western North Pacific tropical cyclones} on the network's own 12-hourly states (ERA5 at 00 and
12\,UTC). \textbf{a}, Seasonal cycle of cyclone centers below 1000\,hPa (20-day mean). \textbf{b}, Cyclones per year by
minimum MSLP. \textbf{c}, \textbf{d}, Density of centers below 994\,hPa.}
\end{figure}

\begin{figure}[!htbp]
\centering
\includegraphics[width=\textwidth]{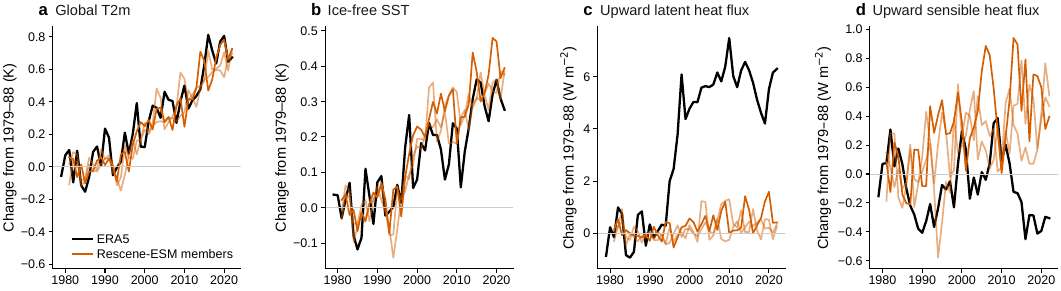}
\caption{\textbf{Historical surface climate, 1979--2022.} Annual changes from 1979--1988 of \textbf{a}, global T2m;
\textbf{b}, ice-free SST; \textbf{c}, \textbf{d}, upward latent and sensible heat fluxes over the ice-free ocean, for ERA5
and the first member of each of the three historical runs.}
\end{figure}

\begin{figure}[!htbp]
\centering
\includegraphics[width=\textwidth]{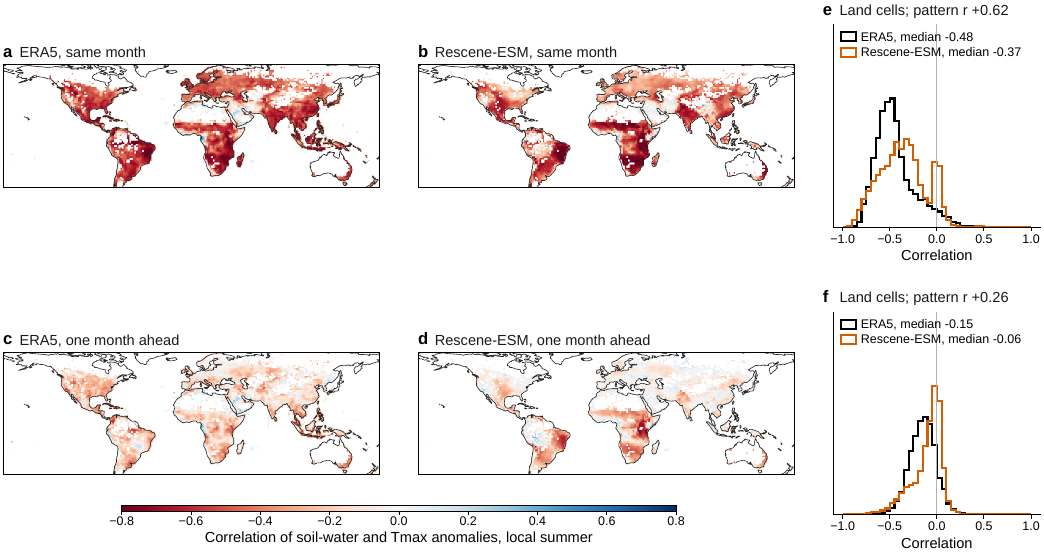}
\caption{\textbf{Land--atmosphere coupling.} Correlation between monthly anomalies of root-zone soil water and of the
daily-maximum temperature in the same month (\textbf{a}, \textbf{b}) and one month later (\textbf{c}, \textbf{d}), local
summer, ERA5 against the historical runs. \textbf{e}, \textbf{f}, Their distributions over land cells.}
\end{figure}

\begin{figure}[!htbp]
\centering
\includegraphics[width=\textwidth]{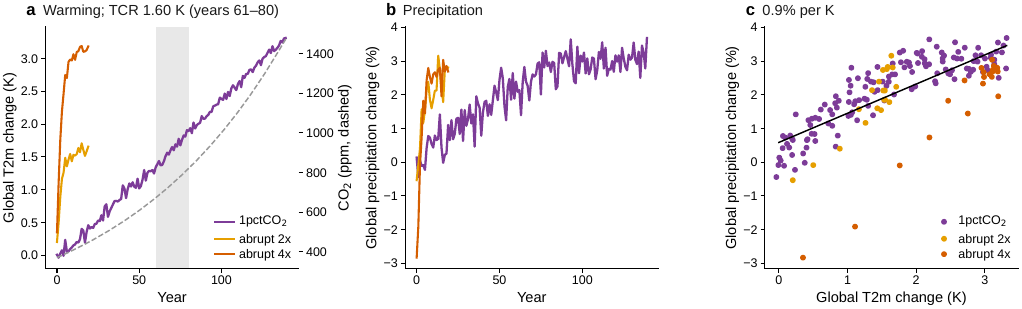}
\caption{\textbf{Transient response.} \textbf{a}, Global T2m change under 1\%-per-year \co\ (with \co, dashed) and the
abrupt experiments; the transient climate response is the mean of years 61--80 (shaded). \textbf{b}, Global
precipitation change. \textbf{c}, Precipitation against temperature change, with the regression of the 1\%-per-year run.}
\end{figure}

\begin{figure}[!htbp]
\centering
\includegraphics[width=\textwidth]{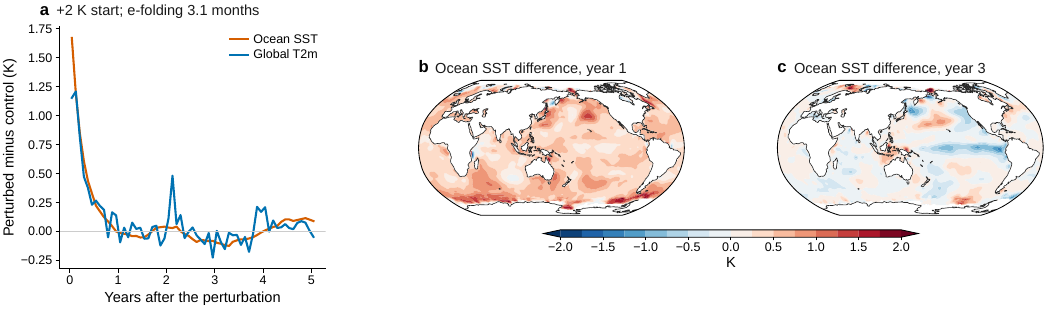}
\caption{\textbf{Return from a warm perturbation.} The control restarted with $+2$\,K open-ocean SST. \textbf{a}, Monthly
difference in open-ocean SST and global T2m from the control; the title gives the e-folding time of the SST difference.
\textbf{b}, \textbf{c}, SST difference in years 1 and 3.}
\end{figure}

\begin{figure}[!htbp]
\centering
\includegraphics[width=\textwidth]{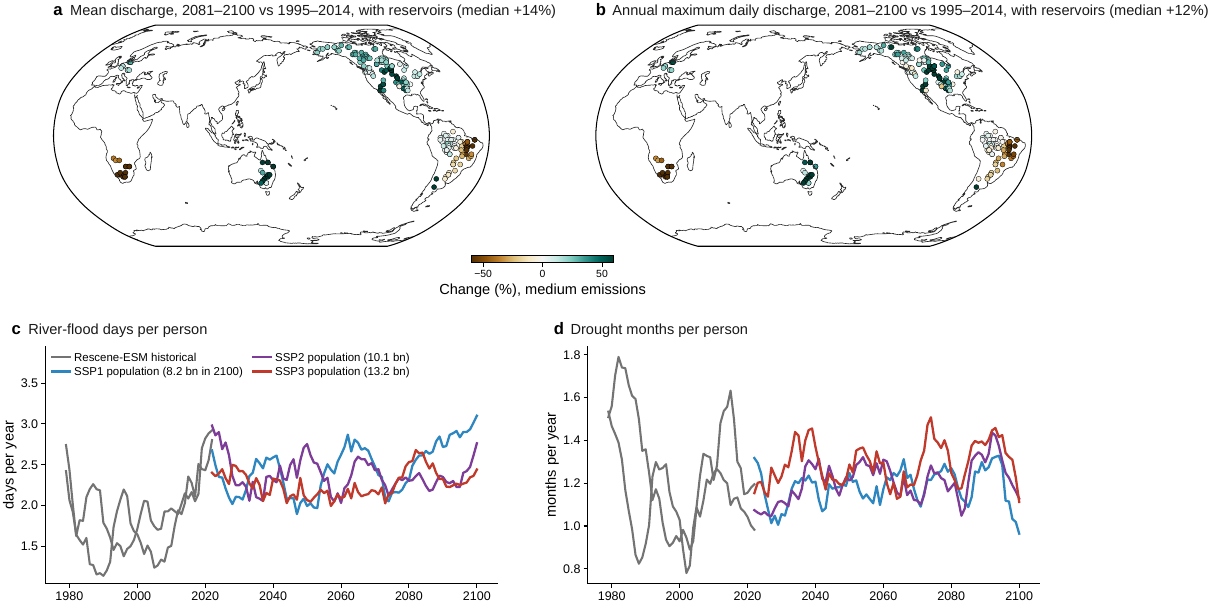}
\caption{\textbf{Future rivers and hazards.} \textbf{a}, Change in mean discharge and \textbf{b}, in the annual maximum
daily discharge at 241 gauges, 2081--2100 against 1995--2014, SSP2 population, with reservoirs. \textbf{c}, River-flood
days and \textbf{d}, drought months per person (five-year running means) in the historical runs and the three population
pathways.}
\end{figure}

\begin{figure}[!htbp]
\centering
\includegraphics[width=\textwidth]{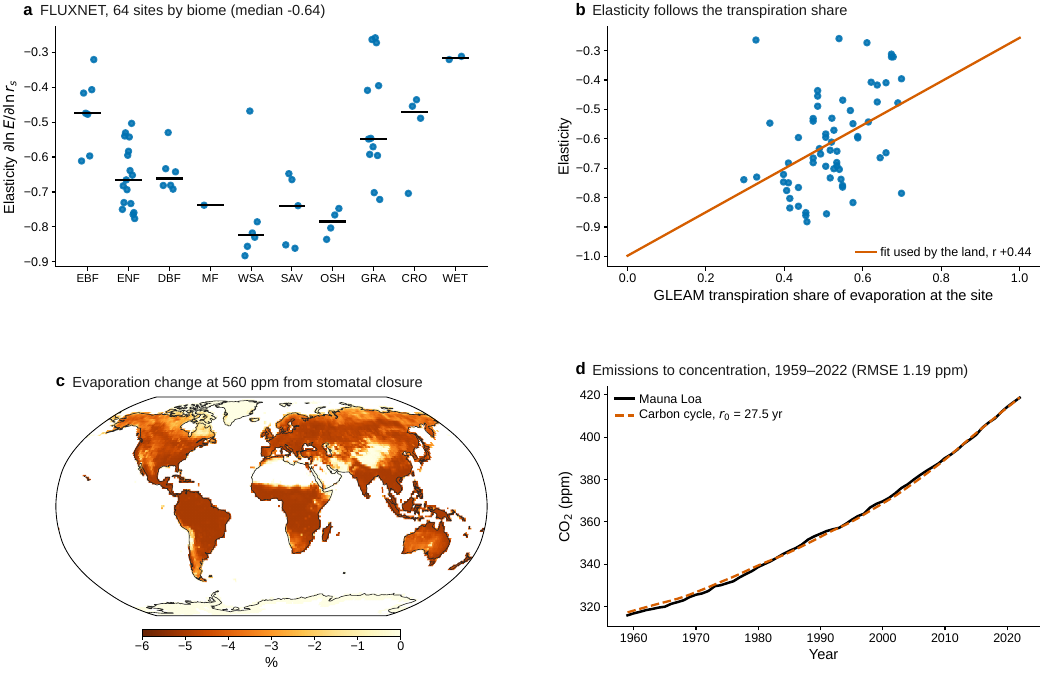}
\caption{\textbf{Component checks against observations.} \textbf{a}, Elasticity of evaporation to stomatal resistance
from Penman--Monteith inversions at 64 FLUXNET sites by biome; bars, medians. \textbf{b}, The elasticity against the
GLEAM transpiration share at each site, with the fit the land uses. \textbf{c}, Change of land evaporation at 560\,ppm
from stomatal closure alone, with the atmosphere held fixed. \textbf{d}, Atmospheric \co\ of the carbon cycle driven by
historical emissions from its 1750 spin-up, with its one fitted constant, against Mauna Loa, 1959--2022.}
\end{figure}

\begin{figure}[!htbp]
\centering
\includegraphics[width=\textwidth]{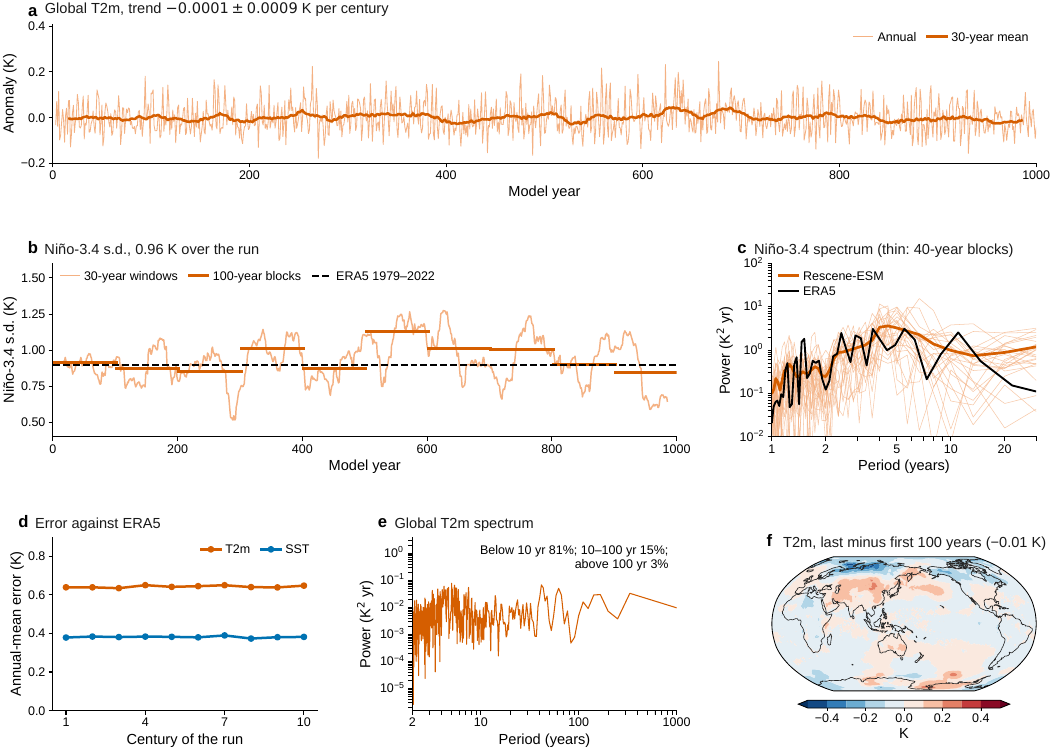}
\caption{\textbf{The 1000-year control} (one member, model years 3--1000). \textbf{a}, Global T2m anomaly, annual and
30-year running mean; the title gives the trend and its standard error, adjusted for lag-1 autocorrelation.
\textbf{b}, Ni\~no-3.4 standard deviation in 30-year windows and 100-year blocks, with ERA5 1979--2022. \textbf{c},
Ni\~no-3.4 power spectrum of 40-year blocks (thin) and their mean (thick), with ERA5 1979--2022. \textbf{d}, Annual-mean
T2m and SST error against ERA5 1979--2018 in each century. \textbf{e}, Power spectrum of annual global T2m, with its
variance by time scale. \textbf{f}, T2m, last 100 years minus the first 100 years.}
\end{figure}

\end{document}